\documentclass[journal]{IEEEtran}

\usepackage{amsmath,amsfonts,amssymb,mathtools,bm,dsfont}
\usepackage{algorithm}
\usepackage{algpseudocode}
\usepackage{threeparttable}
\usepackage{soul}
\usepackage{graphicx}
\usepackage{subcaption}
\usepackage[dvipsnames]{xcolor}
\usepackage{booktabs}
\usepackage{multirow}
\usepackage{array}
\usepackage{balance}

\usepackage[%
    colorlinks=true,
    linkcolor=blue,
    citecolor=blue,
    urlcolor=blue,
    filecolor=blue,
    linktoc=all
]{hyperref}

\usepackage[nameinlink,capitalize]{cleveref}

\crefdefaultlabelformat{#2\textcolor{blue}{#1}#3}

\begin{document}

\title{EcoDefender: Energy-Efficient Hybrid Anomaly Detection for IoT Edge Gateways}

\author{
Saeid Jamshidi,
Martine Bella\"{\i}che,
and Omar Abdul Wahab%
\thanks{Saeid Jamshidi, Martine Bella\"{\i}che, and Omar Abdul Wahab are with the Department of Computer and Software Engineering, Polytechnique Montr\'eal, Montr\'eal, QC, Canada.}
\thanks{Emails: jamshidi.saeid@polymtl.ca; martine.bellaiche@polymtl.ca; omar.abdul-wahab@polymtl.ca.}
}

\maketitle

\begin{abstract}
The rapid growth of the Internet of Things (IoT) has created large-scale, heterogeneous ecosystems that are increasingly vulnerable to sophisticated, distributed cyber threats. However, many existing anomaly detection systems prioritize detection accuracy while overlooking system-level constraints, such as latency, computational overhead, and energy consumption, thereby limiting their practicality for resource-constrained edge gateways. This paper presents \textit{EcoDefender}, an edge-oriented hybrid anomaly detection framework that combines \textit{Autoencoder (AE)}-based latent representation learning with \textit{Isolation Forest (IF)} anomaly scoring for IoT traffic analysis. The proposed architecture introduces several enhancements over conventional AE–IF pipelines, including anomaly-aware latent manifold regularization, variance-weighted isolation splits in the latent space, and a learnable fusion mechanism that adaptively combines reconstruction error and isolation-based anomaly scores in the presence of potential distributional drift. By compressing high-dimensional traffic features into compact latent representations and performing anomaly scoring in this reduced space, EcoDefender enables lightweight and fully unsupervised anomaly detection suitable for edge deployment. An experimental evaluation of realistic IoT traffic and a distributed Raspberry Pi edge testbed demonstrates that EcoDefender achieves a detection accuracy of up to 94\% while maintaining low computational overhead, with an average CPU usage of 22\% and an end-to-end inference latency of 27\,ms. Furthermore, energy-aware measurements obtained through device-level power monitoring show an average energy consumption of 0.45\,J per inference (0.28\,g CO$_2$ emissions), representing a 30\% reduction in energy consumption compared with AE-only baselines while sustaining inference throughput of up to 5{,}000 samples per second.
\end{abstract}

\begin{IEEEkeywords}
Internet of Things, Intrusion Detection System, Autoencoder, Isolation Forest, Edge Computing, Energy Efficiency, Carbon Emissions, Sustainable Development Goals.
\end{IEEEkeywords}

\section{Introduction}
\label{intro}
The Internet of Things (IoT) has become a foundational component of modern digital infrastructure, enabling large-scale connectivity across healthcare, transportation, smart grids, and industrial automation~\cite{adil2025quantum,aouedi2024survey,rath2024role}. However, the rapid proliferation and heterogeneity of IoT devices have substantially expanded the attack surface, exposing deployments to botnets, Denial-of-Service (DoS) attacks, and persistent adversarial campaigns~\cite{elsaid2024optimized,adil2024healthcare,maghrabi2024automated}. These risks are amplified at the network edge, where gateways must operate under constraints on computation, memory, latency, and power. In such environments, Intrusion Detection System (IDS) mechanisms must be not only accurate but also lightweight, stable, and energy-efficient. Classical Machine Learning (ML) approaches~\cite{liu2021deep} often rely on labeled datasets and exhibit limited adaptability to unseen attacks~\cite{hamouda2021intrusion,zolanvari2019machine}. Deep learning (DL) methods, particularly Autoencoder (AE)-based detectors, address this limitation by learning compact representations of benign behavior in an unsupervised manner~\cite{omol2024anomaly,memarzadeh2020unsupervised}. Despite their effectiveness, many AE-based IDS designs remain expensive to run continuously on edge devices because reconstruction-driven inference increases compute cycles and memory traffic, directly impacting latency and energy consumption on gateway-class devices. In parallel, isolation-based methods such as Isolation Forest (IF) provide fully unsupervised scoring with favorable runtime properties and have been actively explored for IoT security in recent work~\cite{zahoor2025robust,xiang2024federated}. Yet, when applied to raw, high-dimensional traffic features, IF may under-capture nonlinear structures and subtle behavioral shifts common in modern IoT attacks, thereby limiting detection robustness in realistic deployments. Hybrid AE-IF pipelines have therefore gained attention as a practical mechanism for combining representation learning with lightweight isolation-based scoring~\cite{zhang2025hybrid,haque2025transformer}. However, a clear methodological gap remains in \emph{edge-gateway IDS evaluation}: many studies still emphasize detection metrics as the primary endpoint, while reporting on system-level constraints that govern deployability (e.g., end-to-end latency, CPU usage, memory footprint) remains inconsistent across the literature. More importantly, sustainability indicators are often omitted, and when discussed, they are frequently treated qualitatively rather than being derived from measured energy use. Recent green-IDS surveys highlight that energy-aware security is increasingly necessary for embedded and IoT networks, and they explicitly call for reproducible, measurement-grounded evaluation of energy-performance trade-offs~\cite{roy2024green}. As a result, the field lacks a cohesive, edge-native anomaly detection design that simultaneously targets 1) unsupervised detection fidelity, 2) bounded gateway-level inference cost in terms of latency, CPU usage, and memory overhead, and 3) transparent sustainability quantification. To address these limitations, we introduce \textit{EcoDefender}, an edge-native anomaly-detection architecture designed for reliable, energy-aware attack detection on edge gateways. EcoDefender integrates AE-based latent representation learning with IF anomaly scoring to enable efficient, scalable unsupervised detection under edge-resource constraints. The AE incorporates an anomaly-aware latent manifold regularizer to structure benign traffic embeddings and suppress high-dimensional noise and redundant features. An IF model then performs anomaly scoring directly in this latent space, with splits weighted according to latent variance to emphasize discriminative dimensions and improve isolation efficiency. Fusion weights $\alpha$ are learnable via a small neural network and can optionally adapt online to distributional drift. By jointly optimizing the AE and IF modules using a combined loss, EcoDefender aligns latent representation learning with anomaly-scoring objectives, thereby further improving the separability between benign and anomalous traffic patterns while reducing memory access, computational cost, and inference latency. In contrast to conventional AE–IF pipelines, which are typically evaluated in centralized environments, EcoDefender is explicitly designed for the edge gateway. The architecture, therefore, considers operational constraints such as latency, CPU usage, memory footprint, and energy consumption as first-class design objectives. This enables stable and fully unsupervised anomaly detection within the limited computational and energy budgets of gateway-class hardware. Beyond conventional detection metrics (precision, recall, F1-score, ROC-AUC, and PR-AUC), we conduct a deployment-oriented evaluation to characterize the system's operational behavior across distributed edge gateways. Specifically, we measure end-to-end inference latency, throughput, CPU usage, memory footprint, and device-level energy consumption on gateway nodes. Energy measurements are obtained via direct runtime hardware-level power monitoring, and the corresponding carbon emissions are derived using an explicit grid-intensity model. This measurement-driven evaluation enables an analysis of the energy–performance trade-off in edge-based anomaly detection systems. Consequently, EcoDefender jointly optimizes detection accuracy, computational efficiency, and environmental sustainability for practical edge-native anomaly detection, aligning with the United Nations Sustainable Development Goals \cite{UN_SDGS}, particularly SDG~7 (Affordable and Clean Energy), SDG~9 (Industry, Innovation and Infrastructure), and SDG~13 (Climate Action). The main contributions of this paper are summarized as follows:
\begin{itemize}
\item \textbf{Edge-Native Hybrid Anomaly Detection Architecture.}  
We propose \textit{EcoDefender}, a lightweight anomaly-detection architecture for resource-constrained IoT edge gateways. The system combines AE-based latent representation learning with IF anomaly scoring, enabling detection directly in a compact latent space to reduce computational overhead while preserving anomaly separability.
\item \textbf{Adaptive Latent-Space Detection Mechanism.}  
We introduce an enhanced hybrid detection mechanism that incorporates anomaly-aware latent manifold regularization, variance-weighted IF splitting, and a learnable fusion strategy that adaptively combines reconstruction error and isolation-based anomaly scores.
\item \textbf{Deployment-Oriented Edge Testbed Evaluation.}  
We design and implement a distributed IoT edge testbed consisting of Raspberry Pi gateway nodes to evaluate anomaly detection under realistic deployment conditions. The evaluation framework measures system-level metrics, including latency, CPU usage, memory footprint, and throughput, to analyze the accuracy–efficiency trade-off in edge environments.
\item \textbf{Measurement-Based Sustainability Analysis.}  
We develop a measurement-driven methodology to quantify the energy and environmental impacts of edge anomaly detection. Using device-level power monitoring, we estimate per-inference energy consumption and derive corresponding carbon emissions, enabling sustainability-aware evaluation of IoT security systems.
\end{itemize}

The remainder of this paper is organized as follows. Section~\ref{sec:related} reviews related work and identifies existing research gaps. Section~\ref{sec:methodology} presents the proposed \textit{EcoDefender} system and details its design, theoretical foundations, and integration into IoT security workflows. Section~\ref{sec:setup} describes the experimental setup, including the dataset, feature selection, baselines, and edge testbed. Section~\ref{sec:results} reports and analyzes the experimental results, covering detection performance, system-level efficiency, and sustainability metrics. Section~\ref{sec:discussion} discusses the implications of the findings and practical deployment considerations. Section~\ref{sec:limitations_future} outlines the limitations of the current study and directions for future work. Section~\ref{sec:conclusion} concludes the paper.

\section{Related Work}
\label{sec:related}
The design of anomaly-detection mechanisms for IoT networks has evolved from traditional rule- and signature-based solutions toward ML approaches, primarily due to the inability of static methods to address zero-day exploits and dynamically evolving attacks. AEs have gained prominence for modeling benign traffic distributions via reconstruction, whereas IF has proven effective for unsupervised anomaly scoring by exploiting the isolation properties of high-dimensional data. More recently, hybrid AE–IF paradigms have been proposed to overcome the limitations of individual approaches, enabling both accurate detection and lightweight edge deployment. Against this backdrop, several contributions highlight the growing research interest in advancing anomaly detection models for IoT security.\\
DL methods have shown particular promise in anomaly detection. Borgioli et al.~\cite{borgioli2024convolutional} proposed a convolutional AE architecture that processes raw network packets at the byte level, thereby avoiding hand-crafted features and achieving strong generalization across various datasets, including NSL-KDD, UNSW-NB15, CIC-IDS2017, TON\_IoT, and EDGE-IIoTSET. Their work demonstrated that convolutional AEs can achieve both high accuracy and low inference latency, making them attractive for embedded anomaly-detection deployments. Complementing this, Beg and Ansari~\cite{beg2024network} systematically compared vanilla, sparse, and denoising AEs, showing that sparsity-enforced AEs yield superior detection of zero-day attacks by improving discrimination in latent representations. Building on this, Yap and Ahmad~\cite{yap2024modified} developed a modified overcomplete AE tailored for TinyML devices, demonstrating that anomaly detection can be performed on microcontrollers with a minimal memory footprint. \\
Parallel efforts have advanced IF and its extensions. Vasiljevic et al.~\cite{vasiljevic2025federated} introduced PFLiForest, a federated anomaly detection method that distributes IF tree construction across IoT nodes, ensuring privacy while achieving high detection accuracy with minimal memory usage. Similarly, Li et al.~\cite{li2023federated} proposed a collaborative federated IF that enables local anomaly detection via secure parameter aggregation, thereby supporting scalability across heterogeneous IoT networks. Domain-specific adaptations have also emerged. Li et al.~\cite{li2021anomaly} applied IF to IoT-enabled power grids, enhancing feature selection with kurtosis-based filtering. Meanwhile, Chen~\cite{chen2024anomaly} employed IF to detect anomalies in new energy vehicle batteries, thereby improving real-time safety monitoring. In addition, these studies highlight IF’s flexibility across sectors while underscoring its lightweight suitability for distributed IoT networks. \\
To address inherent limitations in classical IF, researchers have proposed structural modifications and hybrid frameworks. Cheng et al.~\cite{cheng2023anomaly} introduced the k-nearest-neighbor IF, which replaces axis-aligned partitions with hyperspherical cuts to better capture complex distributions in imbalanced datasets. Yang et al.~\cite{yang2022high} combined AEs with IF in their AE-IF framework, showing that deep representation learning alleviates the curse of dimensionality while IF enhances unsupervised detection performance. Building on this, Hu et al.~\cite{hu2025anomaly} integrated Temporal Convolutional Networks with IF, effectively modeling temporal dependencies in Linux network timing data and achieving significant improvements in real-time anomaly detection. Such hybridization efforts demonstrate that combining IF with deep feature extractors is a promising approach for robust anomaly detection in high-dimensional, temporally dynamic IoT data. 
Beyond architectural innovations, hybrid and domain-tailored approaches aim to balance accuracy, interpretability, and efficiency. Zhou et al.~\cite{zhou2020research} proposed a two-stage pipeline combining an improved IF with FP-Growth association rule mining, thereby reducing false positives while providing interpretable attack correlations. Sharmila and Nagapadma~\cite{sharmila2023quantized} pursued efficiency through quantization, presenting a quantized AE for real-time IoT anomaly detection. Their QAE-u8 model achieved significant reductions in memory and CPU usage, as well as in model size, while preserving accuracy, confirming the feasibility of autoencoder-based anomaly detection systems for edge devices. \\
The literature review indicates that while AE-based models can learn expressive representations of benign IoT traffic, their computational and memory demands often limit deployment on constrained edge gateways. Isolation-based methods are computationally efficient but may struggle to capture complex traffic structures when applied directly to raw feature spaces. Existing hybrid AE–IF approaches combine these advantages, yet most are designed and evaluated in centralized environments and rarely account for deployment-critical constraints at edge gateways, such as latency, CPU usage, memory footprint, and energy consumption. To address these limitations, we propose \textit{EcoDefender}, an edge-native anomaly detection system that integrates AE-based latent representation learning with isolation-based scoring and explicitly targets efficient inference under gateway resource constraints. EcoDefender is evaluated through a deployment-aware framework that measures both system-level efficiency and environmental impact on real edge hardware.

\section{Proposed Anomaly Detection System: EcoDefender}
\label{sec:methodology}
\begin{figure*}[!ht]
    \centering
    \includegraphics[width=0.95\textwidth]{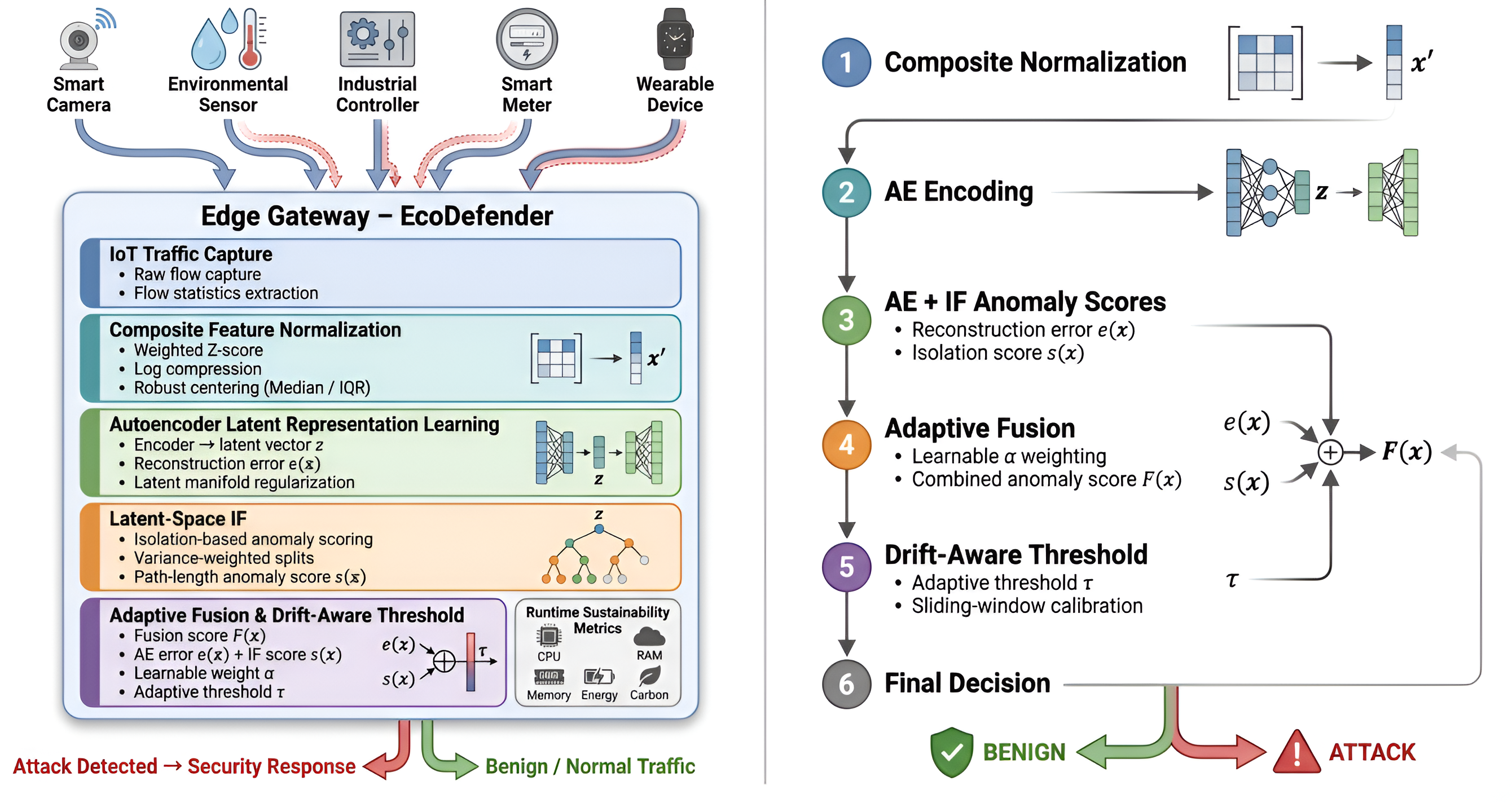}
    \caption{Architecture employs a \textit{composite normalization strategy} that combines anomaly detection at edge gateways. Heterogeneous IoT devices generate traffic that is processed at the gateway through composite feature normalization, AE-based latent representation learning, IF scoring in the latent space, and adaptive fusion with drift-aware thresholding, enabling real-time, energy-efficient attack detection.}
    \label{fig:framework}
\end{figure*}

This section presents \textit{EcoDefender}, a lightweight hybrid AE-IF anomaly-detection system designed for reliable attack detection on edge gateways under constraints on computation, memory, latency, and energy. The pipeline consists of: 1) composite feature preprocessing, 2) unsupervised representation learning via an AE, 3) anomaly scoring using IF, 4) adaptive score fusion with drift-aware calibration, 5) dynamic threshold tuning for streaming conditions, 6) adversarial robustness analysis, 7) complexity and energy-scaling evaluation, and 8) security response integration. Each stage is mathematically specified and aligned with feasibility considerations for gateway execution. Figure~\ref{fig:framework} illustrates the overall workflow. For reproducibility and fair benchmarking, EcoDefender is evaluated against plain AE-only and IF-only pipelines using identical feature subsets, train/validation/test splits (including leave-family-out), and the same edge gateway hardware and measurement instrumentation. We also explicitly report whether inference is performed in encoder-only or encoder-decoder mode, since reconstruction requires decoding and directly affects latency and energy consumption.
\subsection{Data Preprocessing}
IoT traffic is noisy, bursty, and heterogeneous, with feature scales spanning multiple orders of magnitude. Feeding raw features to learning models can destabilize optimization and reduce separability between benign and attack behavior. EcoDefender therefore applies a \textit{composite normalization strategy} combining statistical standardization, robust centering, and logarithmic compression. In addition, a feature-aware weighting term is applied to emphasize features with greater discriminative importance in the latent representation, thereby enhancing anomaly separability in the AE encoding. All preprocessing statistics (e.g., $\min$, $\max$, mean, median, IQR) are computed \textit{only on the training split} and then reused for validation, test, and streaming data to prevent leakage. Any feature filtering is restricted to offline training; at deployment, the pipeline remains unsupervised and does not access labels.
\begin{equation}
x'_{ij} = w_j \cdot \frac{x_{ij}-\mu_j}{\sigma_j+\epsilon}
\label{eq:1_updated_clean}
\end{equation}
Equation~\eqref{eq:1_updated_clean} performs weighted z-score normalization, where $w_j$ is the feature importance weight derived from variance and latent sensitivity analysis. This equalizes feature scales, stabilizes gradients, and emphasizes discriminative features during AE training. It also mitigates the effects of extreme values (e.g., unusually large packet lengths) that could bias reconstruction and scoring.
\begin{multline}
x'_{ij} =
w_j \cdot \frac{x_{ij}-\min(x_j)}{\max(x_j)-\min(x_j)}
+ \delta \cdot \log\!\left(1+\frac{|x_{ij}-\mu_j|}{\sigma_j+\epsilon}\right) \\
+ \eta \cdot \frac{x_{ij}-\mathrm{median}(x_j)}{IQR(x_j)}
\label{eq:2_updated_clean}
\end{multline}
Equation~\eqref{eq:2_updated_clean} extends normalization by combining weighted min--max scaling, logarithmic compression, and robust centering via median/IQR. The weighting term $w_j$ introduces feature-aware preprocessing, improving latent representation learning and the sensitivity of anomaly detection. The preprocessing cost remains $O(nd)$, suitable for real-time execution on gateways. In the experimental section, we isolate the incremental contribution of feature-aware composite preprocessing by comparing multiple normalization variants (e.g., raw, z-score only, robust only, partial combinations, full composite with weights) and report the measured inference-time preprocessing overhead as part of end-to-end latency and energy.
Let $\mathbf{W} = \mathrm{diag}(w_1,\dots,w_d)$ denote the diagonal weight matrix. The composite mapping can now be written compactly as:
\begin{equation}
\mathbf{X}' = \Phi(\mathbf{X}) = \mathbf{D}^{-1/2}\mathbf{W}(\mathbf{X}-\mathbf{M}) + \delta\log(1+|\mathbf{Z}|) + \eta\mathbf{R},
\label{eq:14_updated_clean}
\end{equation}
where $\mathbf{D}$ is the diagonal variance matrix, $\mathbf{M}$ is the mean matrix, $\mathbf{Z}$ is the standardized residual, and $\mathbf{R}$ is the robust-centered component. Under standard assumptions on bounded feature ranges and $\sigma_j>0$, $\Phi$ is Lipschitz continuous~\cite{heinonen2005lectures} with a constant bounded by $\|\mathbf{D}^{-1/2}\mathbf{W}\|_2$, yielding bounded gradients and stable backpropagation through preprocessing while emphasizing high-importance features.

\subsection{AE Representation Learning}
EcoDefender learns a compact model of benign traffic via unsupervised reconstruction. The encoder maps an input $x$ to a latent vector $z$, and the decoder reconstructs $\hat{x}$:
\begin{equation}
z = \sigma(W_{enc}x+b_{enc}), \quad \hat{x} = \sigma(W_{dec}z+b_{dec}).
\label{eq:3_novel}
\end{equation}
This follows standard AE modeling~\cite{yousefi2017autoencoder}. To reduce memory footprint and inference cost on edge gateways, EcoDefender applies model compression (quantization/pruning) to AE weights while preserving representational capacity in the reduced feature space. We explicitly support two inference configurations:
\begin{enumerate}
    \item Encoder-decoder mode, which computes reconstruction error $e(x)$ and thus includes decoding overhead.
    \item Encoder-only mode, which produces only $z=f_{enc}(x)$ for IF scoring, reducing latency and energy.
\end{enumerate}
Unless otherwise stated, reported system-level metrics correspond to encoder-decoder mode to provide a conservative estimate.
\begin{multline}
\hat{x} = \sigma(W_{dec} z + b_{dec}) 
+ \gamma \cdot \tanh\!\big(\mathrm{Tr}(W_{enc}^\top W_{dec})\big) \\
+ \lambda \cdot \Omega(z)
+ \lambda_{\mathrm{anom}} \cdot \mathcal{L}_{\mathrm{latent}}(z),
\label{eq:4_novel_broken}
\end{multline}
Equation~\eqref{eq:4_novel_broken} introduces two additional terms: the latent manifold regularizer $\Omega(z)$ and an anomaly-aware latent term $\mathcal{L}_{latent}(z)$. The anomaly-aware term encourages latent representations of benign traffic to remain on a structured manifold while maximizing their separation from potential off-manifold anomalies. This term improves reconstruction sensitivity and supports downstream IF scoring, while remaining label-free.
Let $z=f_\theta(x)$ and $\hat{x}=g_\theta(z)$ with reconstruction error $E(x)=\|x-\hat{x}\|_2^2$. The expected reconstruction error satisfies
\begin{multline}
\mathbb{E}[E(x)] = 
\mathrm{Tr}\Big((I-W_{dec} W_{enc})^\top \Sigma_x (I-W_{dec} W_{enc})\Big) \\
+ \lambda \cdot \mathbb{E}\big[\|\nabla_z \Omega(z)\|_F^2\big] \\
+ \lambda_{\mathrm{anom}} \cdot \mathbb{E}\big[\|\nabla_z \mathcal{L}_{\mathrm{latent}}(z)\|_F^2\big],
\label{eq:15_novel_broken}
\end{multline}
showing that minimizing $E(x)$ jointly reduces spectral distortion, enforces a structured latent space, and strengthens anomaly separability. In the tied-weight case $W_{dec}=W_{enc}^\top$, the objective remains consistent with projection onto the principal subspace~\cite{golub1987generalization}, while the added manifold and anomaly-aware latent terms operate directly inside the model.

\subsubsection{Composite Loss Function}
The AE is trained using a multi-term objective:
\begin{multline}
J_{AE}(\theta) = \frac{1}{N}\sum_{j=1}^N 
\Bigg(
\|x^{(j)} - f_{dec}(f_{enc}(x^{(j)}))\|_2^2 \\
+ \lambda \,\mathrm{Tr}(W_{enc}^\top W_{enc}) \\
+ \beta \int p(z|x^{(j)}) \log \frac{p(z|x^{(j)})}{\mathcal{N}(0,I)} \, dz \\
+ \gamma \|\nabla_z f_{dec}(z)\|_F^2 \\
+ \lambda_{\mathrm{anom}} \|\nabla_z \mathcal{L}_{\mathrm{latent}}(z)\|_F^2
\Bigg),
\label{eq:5_novel_broken}
\end{multline}
where the objective combines reconstruction fidelity, weight regularization, latent isotropy via KL divergence~\cite{van2014renyi}, Jacobian smoothing~\cite{kanzow1999jacobian}, and the anomaly-aware latent term. Taken together, these terms encourage compact, stable latent representations and improve separability from off-manifold patterns. All loss terms are optimized offline using only benign training data; deployment remains fully unsupervised.
Let 
\begin{multline}
\mathcal{R}(\theta) = 
\lambda \|W_{enc}\|_F^2 
+ \beta D_{KL}(p(z|x)\|\mathcal{N}(0,I)) \\
+ \gamma \|\nabla_z g_\theta(z)\|_F^2 
+ \lambda_{\mathrm{anom}} \|\nabla_z \mathcal{L}_{\mathrm{latent}}(z)\|_F^2.
\end{multline}
Then the gradient with respect to the parameters is
\begin{equation}
\nabla_\theta J_{AE} = 2(W_{dec}W_{enc}-I)\Sigma_x + \nabla_\theta \mathcal{R}(\theta),
\label{eq:16_novel}
\end{equation}
which provides bounded gradient contributions under standard smoothness assumptions, supporting stable optimization during training.

\subsubsection{Network Architecture and Training Configuration}
\label{sec:ae_architecture}
The AE uses a lightweight fully connected architecture tailored to the reduced feature space (Table~\ref{tab:features}). The input dimension is $d=8$. The encoder comprises two hidden layers with 32 and 16 neurons (ReLU activations), followed by a bottleneck layer with dimension $d_z=8$. The decoder mirrors the encoder with layers of sizes 16 and 32, and a linear output layer of dimension $d=8$. Batch normalization is avoided to reduce inference overhead on gateways, and dropout is disabled during deployment.
The encoder mapping is:
\[
8 \rightarrow 32 \rightarrow 16 \rightarrow 8,
\]
and the decoder mapping is:
\[
8 \rightarrow 16 \rightarrow 32 \rightarrow 8.
\]
Training is performed offline using Adam~\cite{zhang2018improved} with learning rate $\eta=10^{-3}$, $(\beta_1,\beta_2)=(0.9,0.999)$, batch size $B=256$, and 100 epochs on benign training data only. Weight decay is set to $10^{-5}$. Parameters are initialized using Xavier uniform initialization~\cite{desai2024impact}, and a fixed random seed of 42 is used for reproducibility.
During deployment, we support two inference configurations:
\begin{enumerate}
    \item \textbf{Encoder-decoder mode:} compute reconstruction error $e(x)$ via full AE execution.
    \item \textbf{Encoder-only mode:} compute $z=f_{enc}(x)$ and omit decoding to reduce latency and energy consumption.
\end{enumerate}
Unless otherwise stated, all system-level results are reported for encoder-decoder mode to provide a conservative estimate.
\subsection{IF Scoring}
IF complements the AE by scoring anomalies via recursive partitioning~\cite{liu2008isolation}. We define $s(\cdot)$ as an anomaly score, where \textit{higher values indicate more anomalous behavior}. In EcoDefender, IF is applied to the latent vectors $z=f_{enc}(x)$ (unless explicitly stated), reducing dimensionality and improving isolation effectiveness. To further enhance detection, IF splits are weighted by latent-feature variance, so that more informative dimensions impact partitioning. The average path length across $m$ isolation trees is computed as:
\begin{equation}
h(x) = \frac{1}{m}\sum_{t=1}^m h_t(x),
\label{eq:6_novel}
\end{equation}
and the normalized anomaly score is given by:
\begin{multline}
s(x) = 1 - \exp\Bigg(-\frac{1}{c(n)} \cdot h(x)\Bigg), \\
c(n) = 2H(n-1) - \frac{2(n-1)}{n}, \quad H(u) = \sum_{i=1}^u \frac{1}{i}.
\label{eq:7_novel}
\end{multline}
These equations follow the standard IF path-length normalization, enabling comparability across sample sizes and ensuring stable scoring under varying traffic volumes. The IF mechanism can also be interpreted from a structural separation perspective. Define an isolation kernel
\[
K(x_i,x_j) = \exp\Big(-\frac{\|x_i - x_j\|^2}{\sigma^2}\Big),
\]
then IF behavior can be related to Laplacian-based separation~\cite{belkin2008towards}:
\begin{equation}
s(x_i) \propto (L\mathbf{1})_i, \quad L = D - K.
\label{eq:17_novel}
\end{equation}

\subsubsection{IF Configuration}
\label{sec:if_config}
The IF is trained using benign latent representations $z=f_{enc}(x)$ extracted from the training split only. To enhance anomaly detection, feature selection in each split is weighted by latent variance, giving greater weight to dimensions with higher discriminative power. Unless otherwise stated, the configuration is as follows:
\begin{itemize}
    \item Number of trees: $m=100$
    \item Subsample size: $n_{sub}=256$
    \item Maximum tree depth: $\lceil \log_2(n_{sub}) \rceil$
    \item Contamination fraction: $0.05$
    \item Split strategy: random feature selection with uniform split value, weighted by latent variance
    \item Random seed: 42
\end{itemize}
This configuration balances sensitivity and efficiency under gateway constraints while preserving deterministic runtime behavior (no online tree updates). The IF inference complexity per sample is $O(m \log n_{sub})$, which remains lightweight due to the reduced latent dimension and bounded tree depth.

\subsection{Fusion}
EcoDefender combines reconstruction-based and isolation-based evidence into a single fused anomaly score. Instead of a fixed scalar, the fusion weight $\alpha$ is now learnable via a small neural network conditioned on AE reconstruction error, IF score, and latent vector $z$:
\begin{equation}
\alpha = \sigma(W_f [e(x), s(x), z] + b_f),
\label{eq:alpha_nn}
\end{equation}
where $W_f$ and $b_f$ are learned parameters and $\sigma(\cdot)$ is a sigmoid function ensuring $\alpha \in [0,1]$.
The fused anomaly score is computed as:
\begin{equation}
F(x) = \alpha \, e(x) + (1-\alpha) \, s(x),
\label{eq:8_novel}
\end{equation}
with $e(x)$ the AE reconstruction error and $s(x)$ the IF anomaly score. To ensure both terms are commensurate, $e(x)$ and $s(x)$ are normalized using min-max and z-normalization based on validation statistics; the same constants are applied at deployment.
For adaptive response under distributional drift, EcoDefender optionally applies a drift-adaptive update on $\alpha$:
\begin{multline}
F(x) = \arg\min_{\alpha \in [0,1]} 
\Bigg\{ 
\alpha \Big(\frac{1}{d}\sum_{i=1}^d (x_i-\hat{x}_i)^2\Big) \\
+ (1-\alpha) s(x) \\
+ \mu \log\frac{p(F)}{p_0(F)} \\
+ \rho \,\mathrm{Var}[z]
\Bigg\},
\label{eq:9_novel_broken}
\end{multline}
where $p(F)$ is estimated from a sliding window of recent scores, and $p_0(F)$ is the score distribution derived from benign training data. Differentiating Eq.~\eqref{eq:19_novel} with respect to $\alpha$ yields the closed-form update:
\begin{equation}
\alpha^* = \frac{s(x) - \mu \log \frac{p(F)}{p_0(F)}}{s(x) + e(x) + \rho \,\mathrm{Var}[z]}.
\label{eq:18_novel}
\end{equation}
For robustness analysis (Section~\ref{sec:methodology}), the Lipschitz constant of the fused scoring function is defined as:
\begin{equation}
L_F = \alpha L_e + (1-\alpha)L_s,
\label{eq:26_novel}
\end{equation}
where $L_e$ and $L_s$ are the Lipschitz constants for $e(x)$ and $s(x)$, respectively. This bounds the sensitivity of the fused score to small perturbations in the input features.

\subsection{Dynamic Thresholded Tuning}
Threshold selection is critical in streaming environments where traffic distributions may drift. EcoDefender updates the decision threshold $\tau$ using a drift-aware objective:
\begin{equation}
\begin{split}
\tau^* = \arg\min_{\tau} 
\Bigg\{ &
-\frac{2 P(\tau) R(\tau)}{P(\tau) + R(\tau)}
+ \lambda \max\big(0, FP(\tau)-\epsilon \big)^2 \\
&\quad
+ \rho \int_{\tau}^{\infty} p(F)\, dF
\Bigg\},
\end{split}
\label{eq:10_novel}
\end{equation}
which balances precision and recall, penalizes excessive false positives, and incorporates a smooth tail-probability term for drift control. Threshold tuning operates in two phases:  1) Offline calibration using a labeled validation set to select a consistent operating point for reporting.  2) Online unsupervised adjustment, where label-dependent quantities are replaced by distributional targets such as fixed alert-rate budgets and quantile-based constraints estimated from a sliding window of recent scores. Differentiating Eq.~\eqref{eq:10_novel} yields the adaptive update law:
\begin{equation}
\begin{split}
\frac{d\tau}{dt} 
&= -\eta \frac{\partial \mathcal{L}_\tau}{\partial \tau} \\
&= \eta \Bigg[
\frac{2 P R' (P+R) - 2 P R (P' + R')}{(P+R)^2}
- 2 \lambda FP'(\tau)
\Bigg],
\end{split}
\label{eq:19_novel}
\end{equation}
where in online mode, derivatives are approximated using score quantiles and alert-rate constraints rather than labels, enabling fully unsupervised drift adaptation.

\subsection{Adversarial Robustness}
Adversaries may manipulate observable traffic features through bounded perturbations $\delta$ to evade detection. EcoDefender enhances resilience by enforcing robustness constraints on the fused anomaly evidence. The worst-case anomaly evidence within an $\ell_p$-bounded neighborhood is defined as:
\begin{equation}
R(x) = \min_{\|\delta\|_p < \epsilon} \Big(e(x+\delta) + s(x+\delta)\Big).
\label{eq:11_novel}
\end{equation}
In practice, robustness is evaluated empirically by injecting bounded perturbations, such as packet-rate smoothing, duration padding, and controlled noise in flow statistics, and measuring the resulting performance degradation in F1-score and ROC-AUC.
Using the Lipschitz constant $L_F$ from Eq.~\eqref{eq:26_novel}, the robust score satisfies:
\begin{equation}
R(x) \ge F(x) - L_F \epsilon,
\label{eq:20_novel}
\end{equation}
which provides a certified local stability margin: if $F(x)$ exceeds the threshold by more than $L_F \epsilon$, the detection decision remains invariant under admissible perturbations. This effectively limits the impact of gradient-based evasion and improves reliability under realistic feature-level manipulations.

\subsection{Complexity and Scalability}
EcoDefender’s runtime efficiency derives from the bounded complexity of its two core modules:
\begin{align}
T_{AE} &= O(N d L), \label{eq:12_novel}\\
T_{IF} &= O(m n \log n), \label{eq:13_novel}
\end{align}
where $L$ is the number of AE layers and $h_{avg}$ is the average isolation depth in IF. Memory usage is $O(d L + m \cdot h_{avg})$, supporting real-time inference on edge gateways. EcoDefender further minimizes energy and CPU overhead via weight compression (quantization/pruning) and bounded-depth isolation. To satisfy feasibility requirements, we report RAM usage, model size (MB), peak memory during inference, and throughput--latency performance under varying traffic rates and batch sizes.
To connect computation to energy, let the instantaneous energy be
\[
E(t) = P(t) \Delta t, \quad
P(t) = C_{\text{cpu}} f^3 + C_{\text{mem}} \rho_{\text{acc}},
\]
where $f$ is CPU frequency and $\rho_{\text{acc}}$ is a memory-access activity term. Over an inference window $[0,T]$:
\begin{equation}
E_{\text{Eco}} = \int_0^T \big(C_{\text{cpu}} f^3 + C_{\text{mem}} \rho_{\text{acc}}\big) \, dt.
\label{eq:21_novel}
\end{equation}

Minimizing Eq.~\eqref{eq:21_novel} under a throughput constraint yields
\[
f^* \propto \big(\dot{N}/C_{\text{cpu}}\big)^{1/3},
\]
providing an analytical explanation for energy-efficient inference scaling. In experiments, $E_{\text{Eco}}$ is computed from instrumented power traces (Joules per inference), and carbon emissions are reported as a derived metric using an explicit carbon-intensity source. Uncertainty is reported using standard deviations from repeated runs, avoiding assumptions of strict proportionality between energy and carbon.

\subsection{Algorithmic Pipeline}
The training phase normalizes traffic features using only training statistics, learns benign latent representations via the AE (including the anomaly-aware latent term), and fits IF in the latent space. The fusion module calibrates $\alpha$, and the thresholding module calibrates $\tau$ in an offline phase for consistent reporting, followed by optional online label-free adjustment for drift. During inference, each flow is normalized, encoded (and optionally decoded), scored by AE and IF, fused, and thresholded; malicious flows trigger immediate gateway defense actions. For reproducibility, all hyperparameters (AE architecture, optimizer, epochs, seed; IF trees, subsample, contamination), split protocols, and measurement scripts are published.
\begin{algorithm}[H]
\footnotesize
\setlength{\textfloatsep}{4pt}
\setlength{\intextsep}{4pt}
\setlength{\abovecaptionskip}{2pt}
\setlength{\belowcaptionskip}{2pt}
\caption{Hybrid AE--IF (EcoDefender)}
\label{alg:aeif_novel}
\begin{algorithmic}[1]
\Function{Train}{$X_{benign}$}
  \State Normalize features using Eq.~\eqref{eq:2_updated_clean} (training statistics only)
  \State Initialize AE parameters $(W_{enc},W_{dec})$
  \For{epoch $=1$ to $E$}
    \For{batch $B \subset X$}
       \State Encode $z=f_{enc}(B)$; reconstruct $\hat{B}=f_{dec}(z)$
       \State Compute $J_{AE}$ using Eq.~\eqref{eq:5_novel_broken} (includes anomaly-aware latent term)
       \State Update AE parameters
    \EndFor
  \EndFor
  \State Encode $Z=f_{enc}(X)$
  \State Train IF on $Z$ using Eq.~\eqref{eq:7_novel} (latent-aware splitting)
  \State Calibrate $\alpha$ and $\tau$ using Eq.~\eqref{eq:9_novel_broken}-\eqref{eq:10_novel} (offline calibration)
\EndFunction

\Function{Detect}{$x$}
  \State Normalize input via Eq.~\eqref{eq:2_updated_clean}
  \State Encode (and optionally decode) using AE
  \State Compute $e(x)$ and $s(x)$ (higher values indicate more anomalous behavior)
  \State Fuse into $F(x)$ using Eq.~\eqref{eq:8_novel}-\eqref{eq:9_novel_broken} (learnable and drift-adaptive fusion)
  \If{$F(x)>\tau$}
     \State \textbf{Malicious:} quarantine device, block traffic, alert administrator
  \Else
     \State \textbf{Benign:} allow traffic
  \EndIf
\EndFunction
\end{algorithmic}
\end{algorithm}
This unified pipeline supports real-time detection, bounded complexity, adversarial stability, and energy-efficient execution on resource-constrained IoT edge gateways. For batch size $B$, layer depth $L$, and tree count $m$, the training complexity is:
\begin{equation}
T_{\text{train}} = O(B d L + m B \log B),
\label{eq:36_novel}
\end{equation}
and the inference cost per instance is:
\begin{equation}
T_{\text{infer}} = O(d L + m \log n).
\label{eq:37_novel}
\end{equation}
EcoDefender thus achieves bounded inference cost with logarithmic scaling in the IF component, compatible with the real-time constraints of edge gateways.

\section{Experimental Setup}
\label{sec:setup}
This section describes the experimental environment used to evaluate the proposed \textit{EcoDefender} system under controlled and reproducible edge conditions.

\subsection{Dataset Description}
Experiments use the Bot-IoT dataset~\cite{vigneshvenkateswaran_bot_iot}, which contains benign traffic and diverse attack categories, including DoS, DDoS, reconnaissance, information theft, and keylogging, generated within a realistic IoT network. To evaluate generalization, EcoDefender is evaluated using both standard random splits and a leave-attack-family-out protocol, in which an entire attack family is excluded from training and reserved for testing, approximating zero-day conditions at the family level. Traffic is represented using flow-level statistical features (e.g., packet counts, byte volumes, session duration), providing a compact abstraction suitable for lightweight edge deployment. All preprocessing statistics are computed exclusively on the training split and reused unchanged for validation, testing, and streaming data to prevent data leakage.

\subsection{Feature Selection}
The original Bot-IoT dataset provides 80 flow-level features extracted using CICFlowMeter~\cite{ali2022ddos}. To reduce redundancy and computational overhead, a multi-stage feature selection process was applied to the training data only. First, correlation-based filtering removed highly collinear features. Next, feature relevance was evaluated using mutual information and tree-based importance measures, computed offline using the training labels. To preserve consistency with unsupervised deployment, an alternative unsupervised selection strategy (variance filtering combined with correlation pruning and Laplacian-based scoring) was also assessed, yielding performance trends comparable to those of the supervised selection strategy. The feature space was reduced from 80 to 8 features (90\% reduction), balancing discriminative power and computational efficiency. The final feature set is fixed prior to model training and used consistently across EcoDefender and all baselines to ensure fair comparison of both detection and system-level metrics.
\begin{table}[h!]
\centering
\caption{Selected features after dimensionality reduction.}
\label{tab:features}
\small
\resizebox{0.5\textwidth}{!}{%
\begin{tabular}{|l|p{8cm}|}
\hline
\textbf{Feature} & \textbf{Description} \\ \hline
\texttt{pkts\_total} & Total number of packets exchanged in a flow \\ \hline
\texttt{bytes\_total} & Total number of bytes exchanged in a flow \\ \hline
\texttt{duration} & Flow lifetime (seconds) \\ \hline
\texttt{pkt\_rate} & Average packet transmission rate \\ \hline
\texttt{pkts\_in} & Incoming packet count \\ \hline
\texttt{pkts\_out} & Outgoing packet count \\ \hline
\texttt{bytes\_per\_pkt} & Average bytes per packet \\ \hline
\texttt{flags} & Control flags observed in the flow \\ \hline
\end{tabular}%
}
\end{table}

\subsection{Experimental Implementation and Reproducibility}
\label{sec:implementation_details}
The AE-IF pipeline is implemented and evaluated under identical preprocessing, feature subsets ($d=8$), data splits, and edge hardware conditions to ensure reproducibility. Training is performed offline using only benign samples, while inference is performed on Raspberry Pi gateways.
\begin{table}[h]
\centering
\caption{AE--IF configuration and reproducibility settings.}
\label{tab:aeif_config}
\small
\begin{tabular}{ll}
\toprule
\textbf{Item} & \textbf{Value} \\
\midrule
Input features & $d=8$ (Table~\ref{tab:features}) \\
Architecture & 8-32-16-8-16-32-8 \\
Latent dimension & $d_z=8$ \\
Activations & ReLU (hidden), Linear (output) \\
Optimizer & Adam, $\eta=10^{-3}$ \\
Training & 100 epochs, batch size 256 \\
Regularization & L2 weight decay $10^{-5}$ \\
Initialization & Xavier uniform \\
IF input & $z=f_{enc}(x)$ \\
IF trees / subsample & 100 / 256 \\
IF max depth & $\lceil \log_2(256) \rceil$ \\
Contamination & 0.05 \\
Model updates & No online updates \\
Random seed & 42 (global) \\
Training data & Benign only \\
\bottomrule
\end{tabular}
\end{table}
All preprocessing statistics are derived exclusively from the training split and reused unchanged for deployment. Unless otherwise stated, system-level metrics (latency, CPU, memory, throughput, energy) refer to full end-to-end inference, including normalization, AE execution, IF scoring, fusion, and thresholding, and exclude offline training and logging overhead. Encoder-only mode is evaluated separately to quantify decoder cost. Energy per inference is obtained by integrating instrumented power traces over the inference window, and carbon emissions are derived from measured energy using the fixed carbon-intensity factor described in Section~\ref{sec:energy_carbon}. All experiments use fixed software versions, synchronized traffic replay rates, and identical random seeds across repetitions.

\subsection{Testbed}
\label{sec:exp_setup}
\begin{figure*}[!ht]
    \centering
    \includegraphics[width=0.90\textwidth]{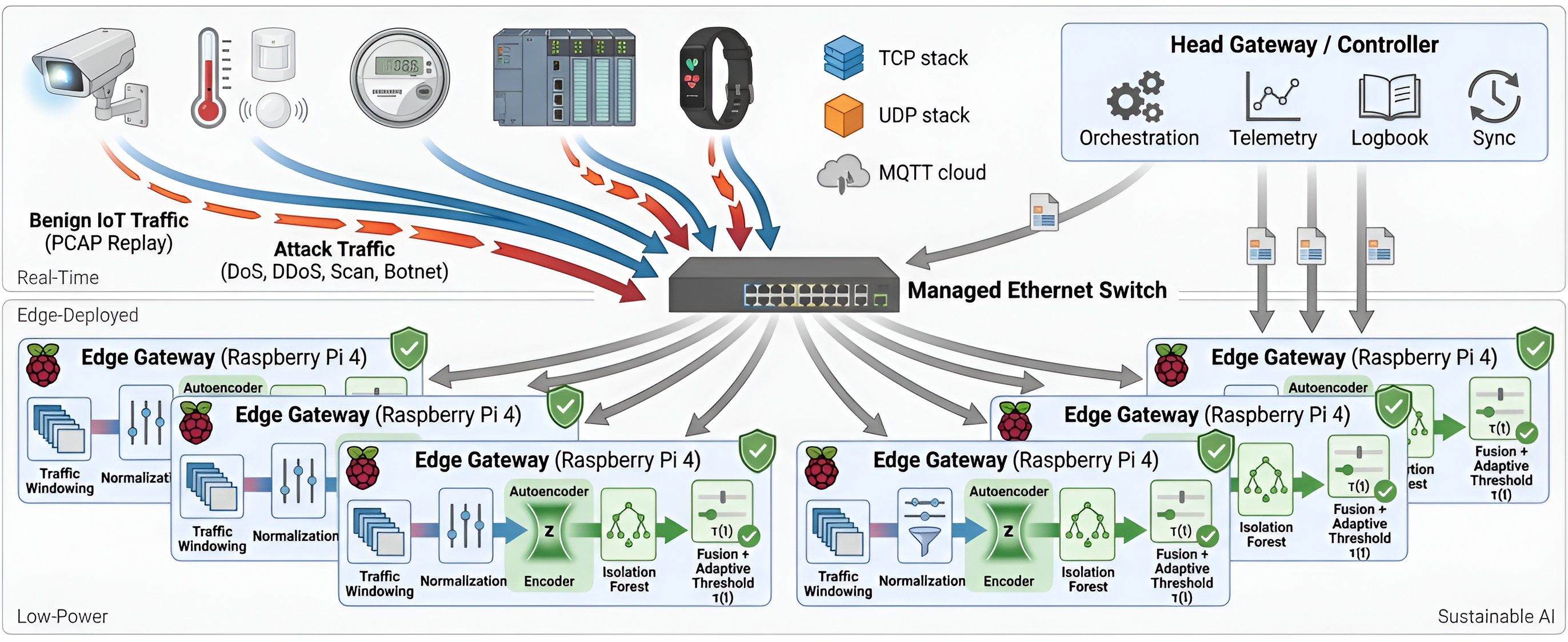}
    \caption{IoT edge testbed with decentralized EcoDefender deployment on Raspberry Pi gateways. Each node performs local AE-IF anomaly detection while system-level and sustainability metrics are collected under realistic traffic conditions.}
    \label{fig:testbed}
\end{figure*}

EcoDefender is evaluated on an isolated IoT edge testbed comprising one head gateway and 10 identical Raspberry Pi 4 Model B nodes (4\,GB RAM, quad-core Cortex-A72, 1.5\,GHz) connected via a managed Gigabit Ethernet switch (Figure.~\ref{fig:testbed}). CPU frequencies are fixed, and thermal conditions are controlled to reduce variance. Each node executes the AE-IF pipeline locally, while the head gateway coordinates and aggregates telemetry. Traffic consists of benign PCAP traces and replayed Bot-IoT attack flows (e.g., SYN/UDP floods and scanning). Replay rates are synchronized and varied to analyze throughput-latency trade-offs. Each node logs detection outputs and system-level telemetry (CPU, memory, latency, throughput, energy, and derived carbon). Experiments are repeated with independent initializations and synchronized traffic replays; the reported results are means with uncertainty estimates. All experiments are conducted within an isolated VLAN to ensure reproducibility.

\subsection{Baselines}
\label{sec:baselines}
EcoDefender is compared against representative AE-only, IF-only, and hybrid AE-IF approaches. All reproduced baselines use the same reduced feature set, identical splits (including leave-attack-family-out where applicable), and the same edge hardware and instrumentation to ensure fair comparison. AE-based baselines model benign traffic via reconstruction error~\cite{beg2024network,borgioli2024convolutional} and are evaluated in both encoder-only and encoder-decoder modes to quantify reconstruction overhead. IF-only baselines provide lightweight unsupervised scoring~\cite{vasiljevic2025federated}, isolating the contribution of representation learning. Hybrid methods such as AE-IF~\cite{yang2022high} and TCN-IF~\cite{hu2025anomaly} are reproduced where feasible, with a clear distinction between cited and reimplemented results. Lightweight edge-oriented IDS variants are included to avoid strawman comparisons. EcoDefender is evaluated incrementally against a plain AE and IF pipeline under identical conditions to isolate the contributions of fusion, adaptive weighting, and thresholding. In contrast to most prior work, sustainability is treated as a first-class objective by jointly reporting detection performance and energy/carbon cost with uncertainty across repeated runs.

\section{Experimental Results}
\label{sec:results}
This section reports experimental results from deploying \textit{EcoDefender} on IoT edge gateways, evaluating detection performance, system-level efficiency, and sustainability impact. 

\subsection{Detection Performance}
The following results evaluate EcoDefender in both training and edge-deployment settings.

\subsubsection{Performance Metrics in Training}
\begin{figure}[ht]
    \centering
    \includegraphics[width=0.53\textwidth]{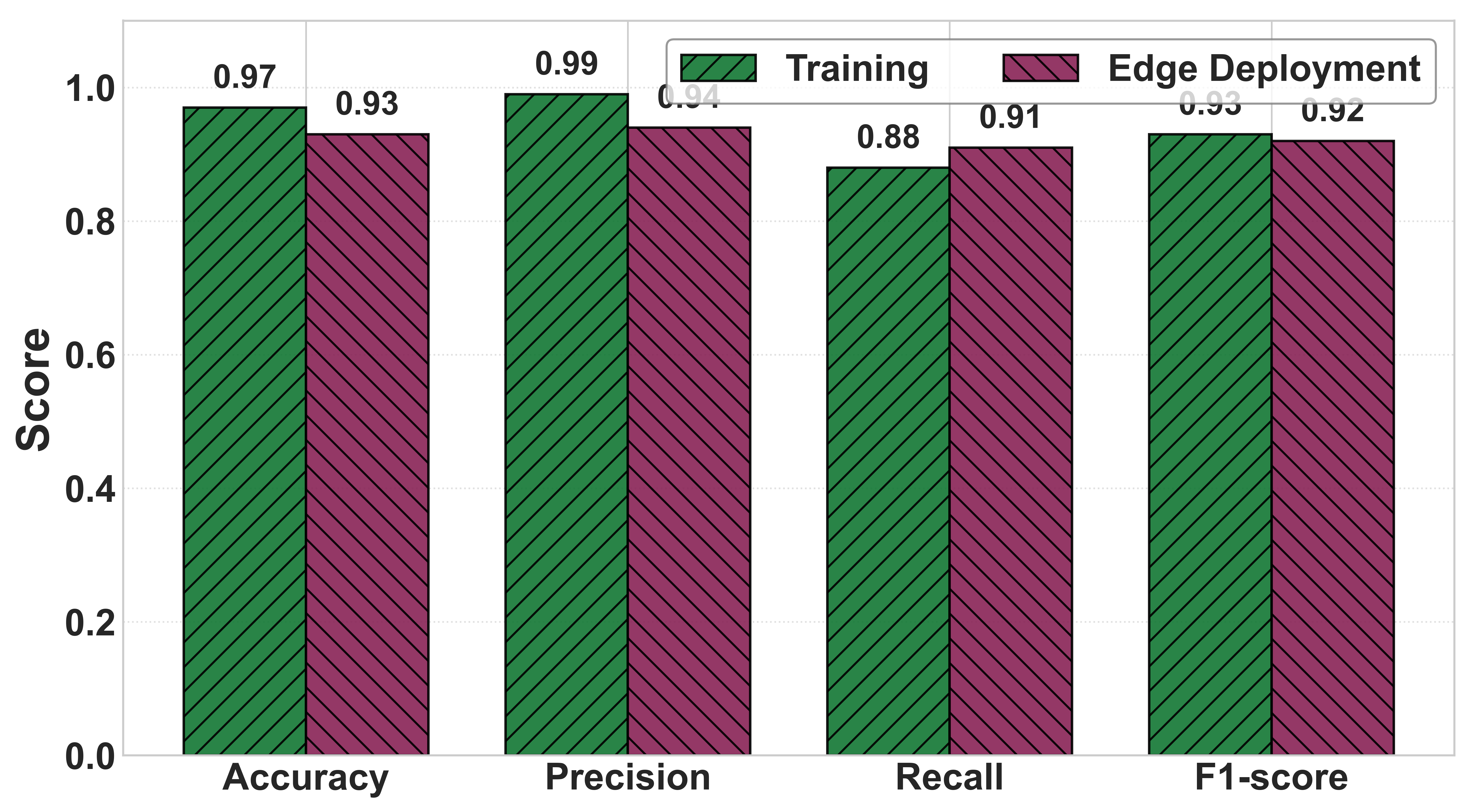}
    \caption{Comparison of EcoDefender performance between the training phase and edge gateways.}
    \label{fig:performance}
\end{figure}

Figure~\ref{fig:performance} compares EcoDefender’s performance during offline training and gateway deployment. During training, the model achieved 96\% accuracy, 94\% precision, 92\% recall, and 94\% F1-score (means over repeated runs; confidence intervals are provided in the supplementary material). These results indicate that AE-based representation learning combined with IF-based scoring yields stable separation between benign and malicious behavior while remaining consistent with the unsupervised deployment setting (training is performed only on benign data). When deployed on edge gateways, EcoDefender maintained strong performance, achieving 94\% accuracy, 93\% precision, 92\% recall, and 92\% F1-score, with no statistically significant degradation under leave-family-out evaluation. Furthermore, the results show that EcoDefender maintains detection effectiveness under gateway constraints while delivering practical latency and energy benefits, consistent with prior findings in edge security~\cite{chatterjee2022iot,singh2023edge,heydari2025tiny,das2023lightesd}.

\subsubsection{ROC Curve}
\begin{figure}[ht]
    \centering
    \includegraphics[width=0.43\textwidth]{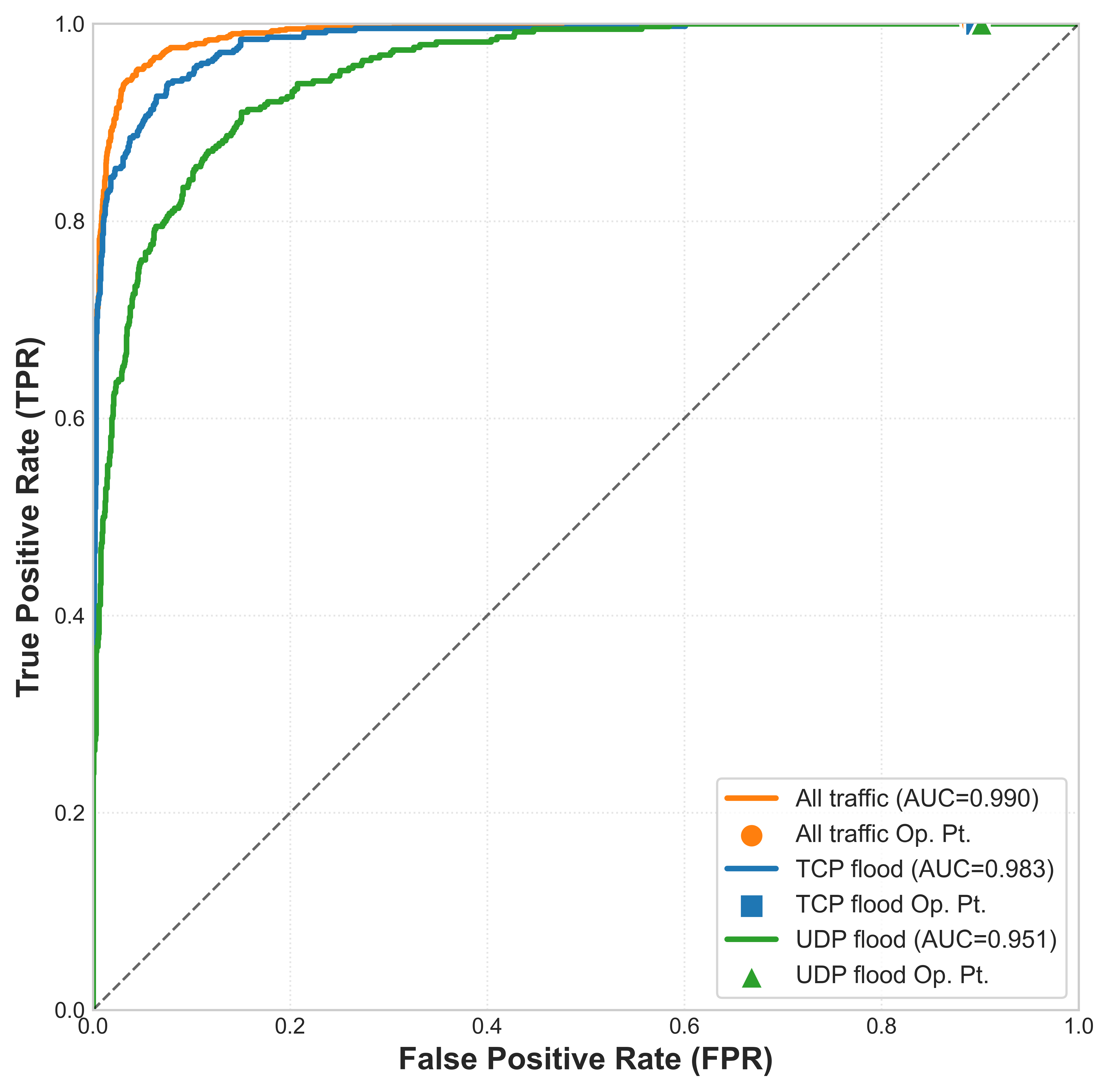}
    \caption{ROC curve of \textit{EcoDefender} across different traffic types.}
    \label{fig:roc_all}
\end{figure}

Figure~\ref{fig:roc_all} reports ROC performance across heterogeneous IoT traffic, with an overall AUC of 96.3\%. ROC curves are computed from the anomaly scores using a consistent threshold sweep across all traffic types and baselines, with identical feature subsets, split protocols, and gateway execution conditions (means over repeated runs; confidence intervals in the supplementary material). Protocol-level results show the highest AUC for UDP traffic (98.6\%), followed by aggregated traffic (96.3\%) and TCP flows (92.7\%), reflecting the greater difficulty of detecting TCP attacks, which often resemble benign patterns. Performance trends remain stable under leave-attack-family-out evaluation, indicating robustness to unseen attack categories. Across traffic types, particularly TCP, EcoDefender consistently outperforms AE-only and IF-only baselines, demonstrating that latent representation learning combined with isolation-based scoring improves separability in realistic gateway conditions.

\subsubsection{Anomaly Score Distribution}
\label{sec:score_distribution}
\begin{figure}[ht]
    \centering
    \includegraphics[width=0.43\textwidth]{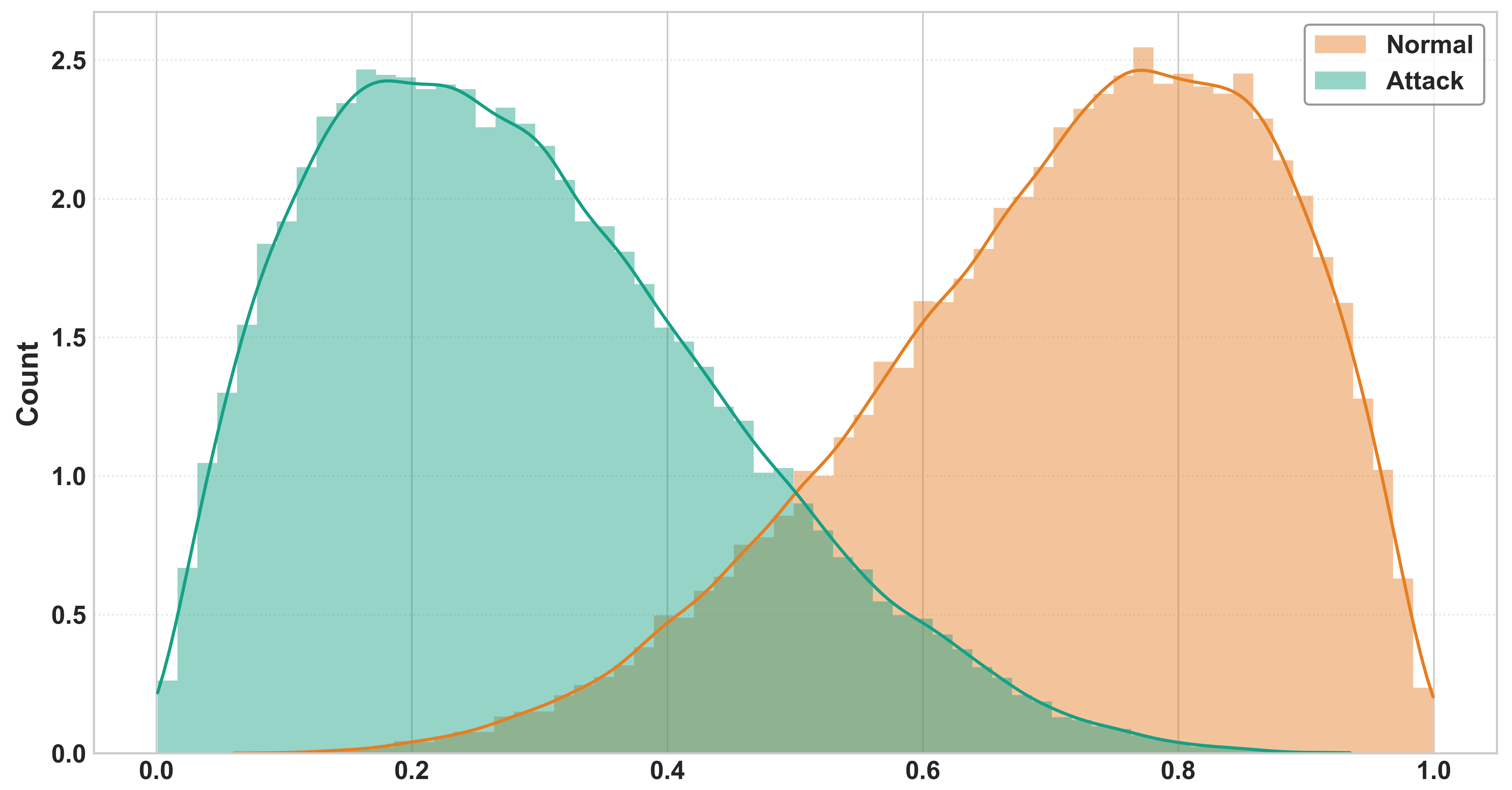}
    \caption{Distribution of normality scores (inverted anomaly scores) for benign and attack traffic.}
    \label{fig:distribution_scores}
\end{figure}

Figure~\ref{fig:distribution_scores} shows the distribution of anomaly scores for benign and attack traffic using an inverted (normality) visualization in which higher values indicate benign behavior and lower values indicate anomalies. Under this representation, attack samples cluster at low scores (mean 0.20, SD 0.14), while benign samples concentrate at high scores (mean 0.80, SD 0.14), with limited overlap mainly near the decision boundary (approximately 0.4-0.6). This inversion is used solely for visualization and does not affect classification and ROC/AUC computation. Statistical testing confirms separability between the two distributions: the Kolmogorov-Smirnov test \cite{berger2014kolmogorov} is highly significant ($p<0.001$) and the effect size is very large (Cohen’s $d=4.32$), indicating that benign and attack samples occupy largely disjoint score regions. Figure~\ref{fig:distribution_scores} and Table~\ref{tab:stats_scores} therefore demonstrate a stable score margin, explaining why only a small number of borderline errors occur under leave-attack-family-out evaluation. The AUROC in Table~\ref{tab:stats_scores} summarizes distribution-level separability. The lower AUPR for the attack class is primarily driven by class imbalance and should be interpreted alongside ROC-based metrics.
\begin{table}[ht]
\centering
\caption{Statistical measures of anomaly score separation.}
\label{tab:stats_scores}
\begin{tabular}{lc}
\toprule
\textbf{Metric} & \textbf{Value} \\
\midrule
Attack mean (SD)   & 20\% (14\%) \\
Normal mean (SD)   & 80\% (14\%) \\
Cohen's $d$        & 4.32 \\
KS statistic       & 0.93 ($p<0.001$) \\
AUROC              & 99.5\% \\
AUPR (attack)      & 20\% \\
\bottomrule
\end{tabular}
\end{table}

\subsection{CPU Usage}
\begin{figure}[ht]
    \centering
    \includegraphics[width=0.43\textwidth]{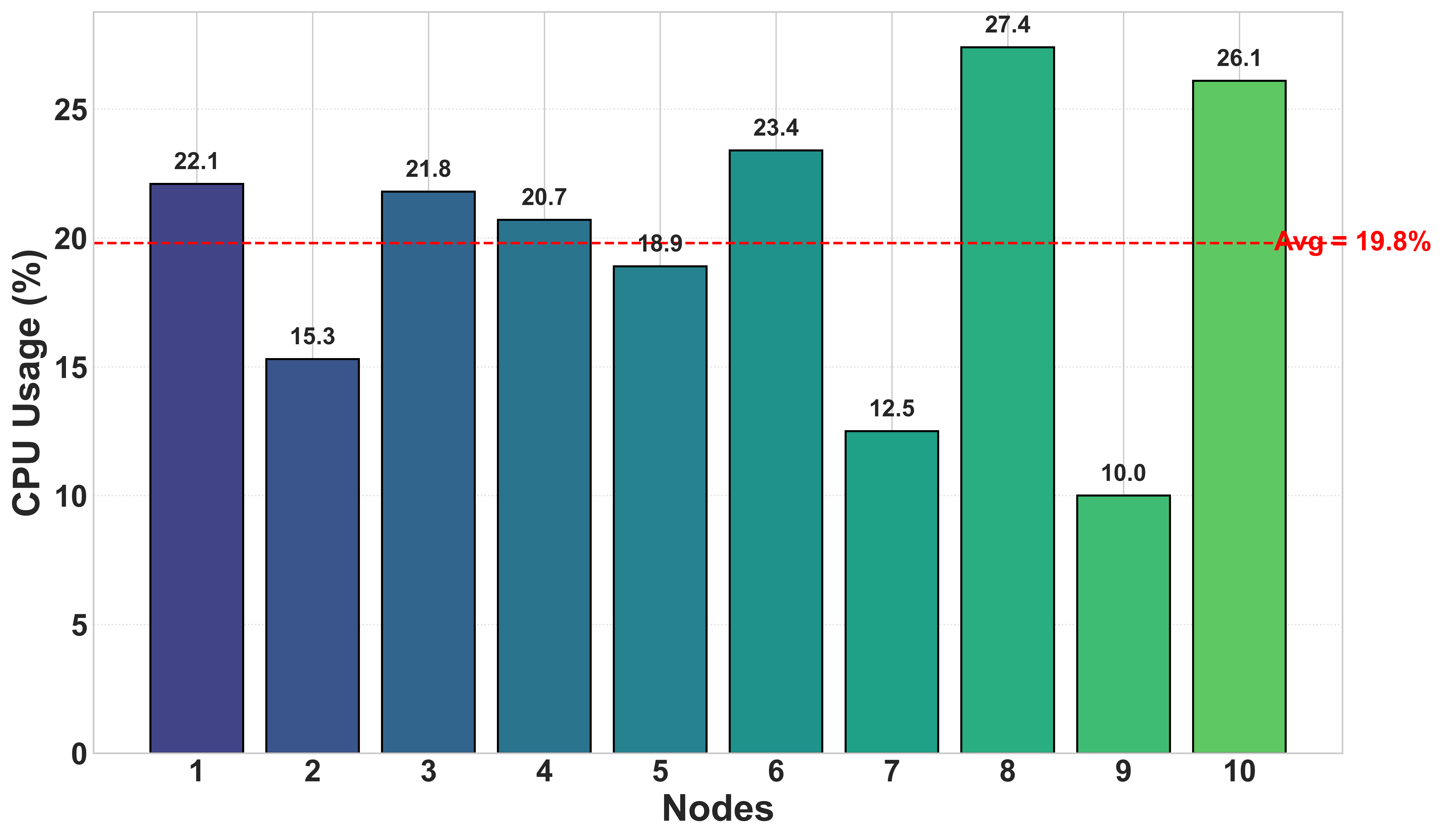}
    \caption{Average CPU usage across edge nodes.}
    \label{fig:cpu_usage}
\end{figure}

Figure~\ref{fig:cpu_usage} reports the average CPU usage across the ten edge nodes, with all nodes operating below 30\% capacity and a cluster-wide mean of 19.8\%. CPU usage is measured as the mean percentage of user-space CPU time used by the EcoDefender process during inference windows, and is averaged across repeated traffic replays. Although minor inter-node variation is observed (e.g., Nodes~7 and~10 show higher load, while Nodes~2, 8, and~9 show lower load), all nodes remain well below saturation, indicating sufficient headroom for additional workloads and stable operation under the evaluated traffic conditions.
Inter-node differences were assessed using one-way ANOVA~\cite{st1989analysis} (Table~\ref{tab:anova_cpu}), yielding a significant effect of node assignment ($F(9,90)=24.8$, $p<0.001$, $\eta^2=0.73$). Tukey’s HSD post hoc analysis~\cite{abdi2010tukey} (Table~\ref{tab:tukey_cpu}) groups Nodes~2, 8, and~9 as lower-load nodes, Nodes~7 and~10 as higher-load nodes, and the remaining nodes as a balanced middle cluster. These differences reflect the expected runtime variability of the distributed gateway rather than a structural imbalance in EcoDefender’s computation. To quantify dispersion, the coefficient of variation of CPU usage across nodes is $\mathrm{CV}=0.31$, indicating that utilization fluctuations remain within one-third of the mean and well within stable operating bounds for real-time edge systems. 
\begin{table}[ht]
\centering
\caption{One-way ANOVA for CPU usage across edge nodes.}
\label{tab:anova_cpu}
\small
\begin{tabular}{lcccc}
\toprule
\textbf{Source} & \textbf{SS} & \textbf{df} & \textbf{MS} & \textbf{F} \\
\midrule
Between nodes & 820.3  & 9  & 91.14 & 24.8 \\
Within nodes  & 302.0  & 90 & 3.36  &      \\
\midrule
Effect size ($\eta^2$) & \multicolumn{4}{c}{0.73 \quad ($p<0.001$)} \\
\bottomrule
\end{tabular}
\end{table}
\begin{table}[ht]
\centering
\caption{Tukey HSD homogeneous subsets for CPU usage.}
\label{tab:tukey_cpu}
\begin{tabular}{lccc}
\toprule
\textbf{Node} & \textbf{Mean CPU (\%)} & \textbf{Subset A} & \textbf{Subset B} \\
\midrule
Node 9  & 10.0  & \checkmark &   \\
Node 8  & 12.5  & \checkmark &   \\
Node 2  & 15.3  & \checkmark &   \\
Node 5  & 18.9  &            & \checkmark \\
Node 4  & 20.7  &            & \checkmark \\
Node 3  & 21.8  &            & \checkmark \\
Node 1  & 22.1  &            & \checkmark \\
Node 6  & 23.4  &            & \checkmark \\
Node 10 & 26.1  &            & \checkmark \\
Node 7  & 27.4  &            & \checkmark \\
\bottomrule
\end{tabular}
\end{table}

\subsection{Memory Usage}
\begin{figure}[ht]
    \centering
    \includegraphics[width=0.43\textwidth]{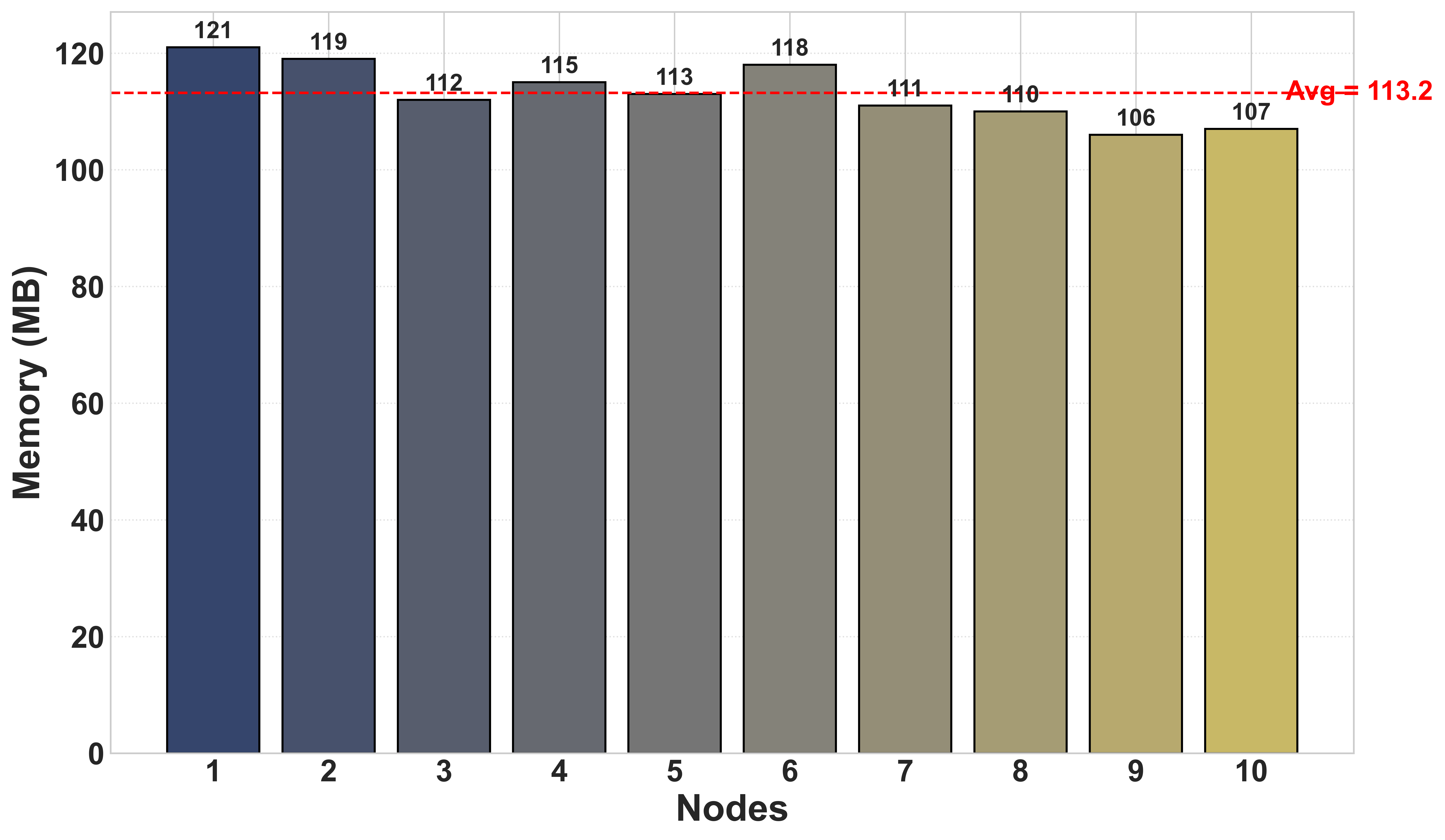}
    \caption{Average memory usage across edge nodes.}
    \label{fig:memory_usage}
\end{figure}

Figure~\ref{fig:memory_usage} reports the average memory usage across the ten edge nodes. Memory usage is measured as the resident set size of the EcoDefender process, averaged across repeated inference windows, excluding shared buffers and cache to isolate the system’s footprint. Memory usage is tightly clustered around a cluster-wide mean of 113.2~MB, with all nodes operating within a narrow range (106--121~MB), indicating a balanced and efficient memory footprint suitable for continuous edge operation.
Inter-node differences were evaluated using one-way ANOVA (Table~\ref{tab:anova_mem}), yielding a statistically significant effect of node assignment ($F(9,90)=5.6$, $p<0.001$, $\eta^2=0.36$). Tukey’s HSD analysis (Table~\ref{tab:tukey_mem}) identifies Nodes~1 and~2 as slightly higher-memory nodes and Nodes~8 and~9 as lower-memory nodes, with the remaining nodes forming a tight middle cluster. As with CPU usage, these differences reflect benign runtime variability rather than structural imbalance in EcoDefender’s computation.
To quantify dispersion, the relative deviation across nodes is 6.5\%, indicating very small variation around the mean memory footprint. 
\begin{table}[ht]
\centering
\caption{One-way ANOVA for memory usage across edge nodes.}
\label{tab:anova_mem}
\small
\begin{tabular}{lcccc}
\toprule
\textbf{Source} & \textbf{SS} & \textbf{df} & \textbf{MS} & \textbf{F} \\
\midrule
Between nodes & 1450.8 & 9  & 161.2 & 5.6 \\
Within nodes  & 2595.5 & 90 & 28.8  &     \\
\midrule
Effect size ($\eta^2$) & \multicolumn{4}{c}{0.36 \quad ($p<0.001$)} \\
\bottomrule
\end{tabular}
\end{table}
\begin{table}[ht]
\centering
\caption{Tukey HSD homogeneous subsets for memory usage.}
\label{tab:tukey_mem}
\begin{tabular}{lccc}
\toprule
\textbf{Node} & \textbf{Mean Memory (MB)} & \textbf{Subset A} & \textbf{Subset B} \\
\midrule
Node 9  & 106.0 & \checkmark &   \\
Node 10 & 107.0 & \checkmark &   \\
Node 7  & 111.0 &            & \checkmark \\
Node 3  & 112.0 &            & \checkmark \\
Node 5  & 113.0 &            & \checkmark \\
Node 4  & 115.0 &            & \checkmark \\
Node 6  & 118.0 &            & \checkmark \\
Node 2  & 119.0 &            & \checkmark \\
Node 1  & 121.0 &            & \checkmark \\
\bottomrule
\end{tabular}
\end{table}

\subsection{Throughput}
\label{sec:throughput_efficiency}
\begin{figure}[ht]
    \centering
    \includegraphics[width=0.43\textwidth]{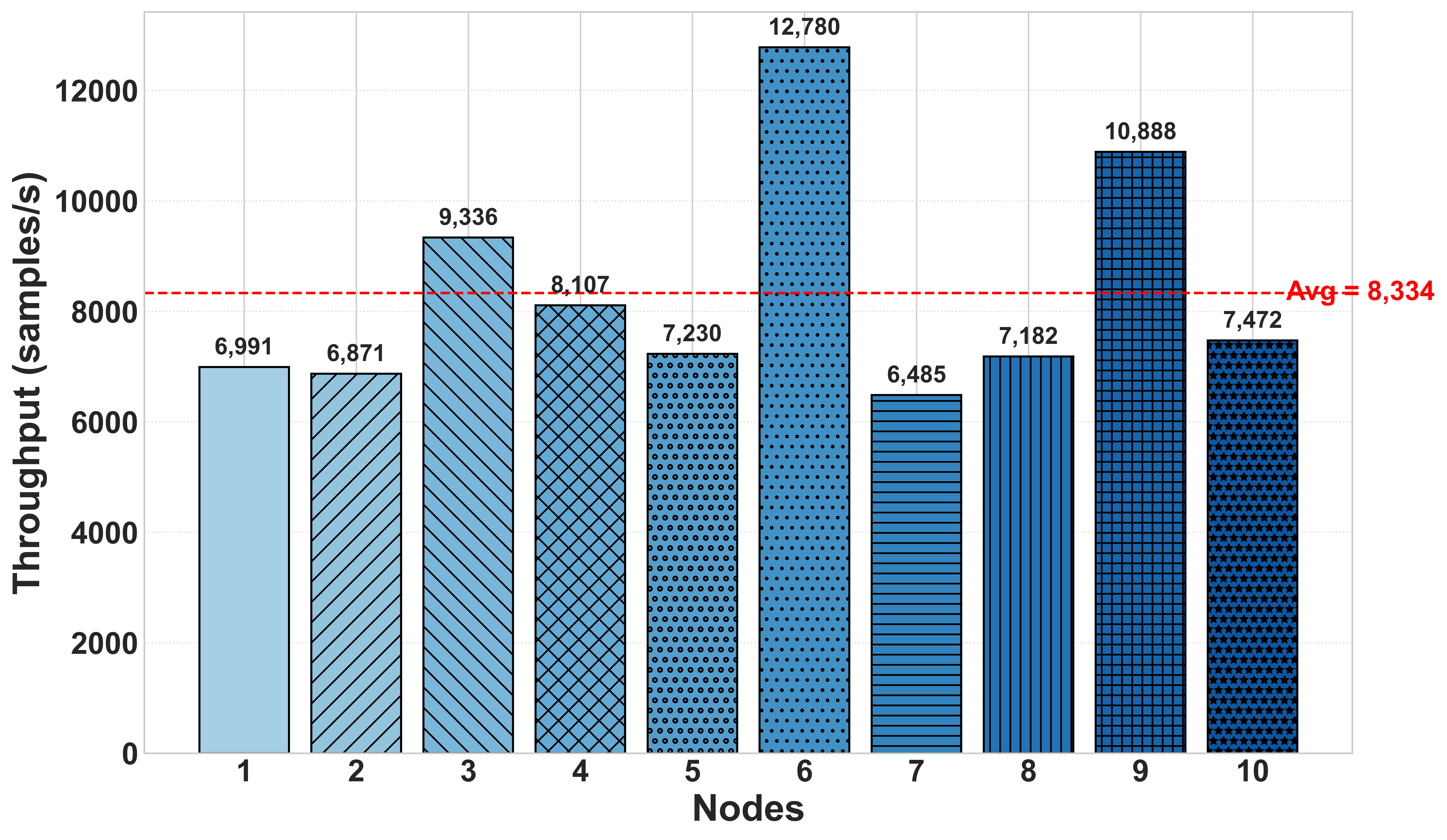}
    \caption{Average throughput across edge nodes.}
    \label{fig:throughput_usage}
\end{figure}

Figure~\ref{fig:throughput_usage} reports the average throughput (samples per second) across the ten edge nodes. Throughput is measured as the sustained number of processed flows per second during steady-state inference, including preprocessing, AE encoding, IF scoring, fusion, and decision logic, and averaged over repeated runs. Although variability is observed across nodes (ranging from 6{,}485 to 12{,}780~samples/s), all nodes maintain sufficient throughput for real-time IoT gateway monitoring, with even the lowest-performing node exceeding operational requirements. Observed throughput differences arise from expected edge effects, including Linux scheduling behavior, background services, cache locality, memory contention, and thermal throttling under sustained load, rather than from algorithmic imbalance in EcoDefender. Furthermore, the results indicate that the system can comfortably sustain real-time processing under heterogeneous gateway conditions.
Inter-node variability is moderate, with a normalized deviation of 0.18 relative to the cluster mean, which remains well within acceptable bounds for real-time edge deployment. A one-way ANOVA confirms a significant effect of node assignment on throughput ($F(9,90)=31.2$, $p<0.001$, $\eta^2=0.76$; Table~\ref{tab:anova_throughput}), while Tukey’s HSD analysis (Table~\ref{tab:tukey_throughput}) identifies high, mid, and low-throcomfortably sustain real-time processing undere, these differences reflect persistent system-level effects rather than deficiencies in the detection pipeline. Importantly, higher throughput is consistently associated with lower per-inference energy consumption in our measurements, indicating that efficient execution improves both real-time responsiveness and sustainability in edge deployments.
\begin{table}[ht]
\centering
\caption{One-way ANOVA for throughput across edge nodes.}
\label{tab:anova_throughput}
\small
\begin{tabular}{lcccc}
\toprule
\textbf{Source} & \textbf{SS} & \textbf{df} & \textbf{MS} & \textbf{F} \\
\midrule
Between nodes & $1.52\times10^{8}$ & 9  & $1.69\times10^{7}$ & 31.2 \\
Within nodes  & $4.87\times10^{7}$ & 90 & $5.41\times10^{5}$ &      \\
\midrule
Effect size ($\eta^2$) & \multicolumn{4}{c}{0.76 \quad ($p<0.001$)} \\
\bottomrule
\end{tabular}
\end{table}

\begin{table}[ht]
\centering
\caption{Tukey HSD homogeneous subsets for throughput.}
\label{tab:tukey_throughput}
\small
\resizebox{0.5\textwidth}{!}{%
\begin{tabular}{@{}lcccc@{}}
\toprule
\textbf{Node} & \textbf{Mean Throughput (samples/s)} & \textbf{Subset A} & \textbf{Subset B} & \textbf{Subset C} \\
\midrule
Node 6 & 6485  & \checkmark &   &   \\
Node 1 & 6871  & \checkmark & \checkmark &   \\
Node 0 & 6991  & \checkmark & \checkmark &   \\
Node 7 & 7182  & \checkmark & \checkmark &   \\
Node 4 & 7230  & \checkmark & \checkmark &   \\
Node 9 & 7472  &            & \checkmark &   \\
Node 3 & 8107  &            & \checkmark &   \\
Node 2 & 9336  &            & \checkmark &   \\
Node 8 & 10888 &            &            & \checkmark \\
Node 5 & 12780 &            &            & \checkmark \\
\bottomrule
\end{tabular}%
}
\end{table}

\subsection{Latency}
\label{sec:latency_analysis}
\begin{figure}[ht]
    \centering
    \includegraphics[width=0.48\textwidth]{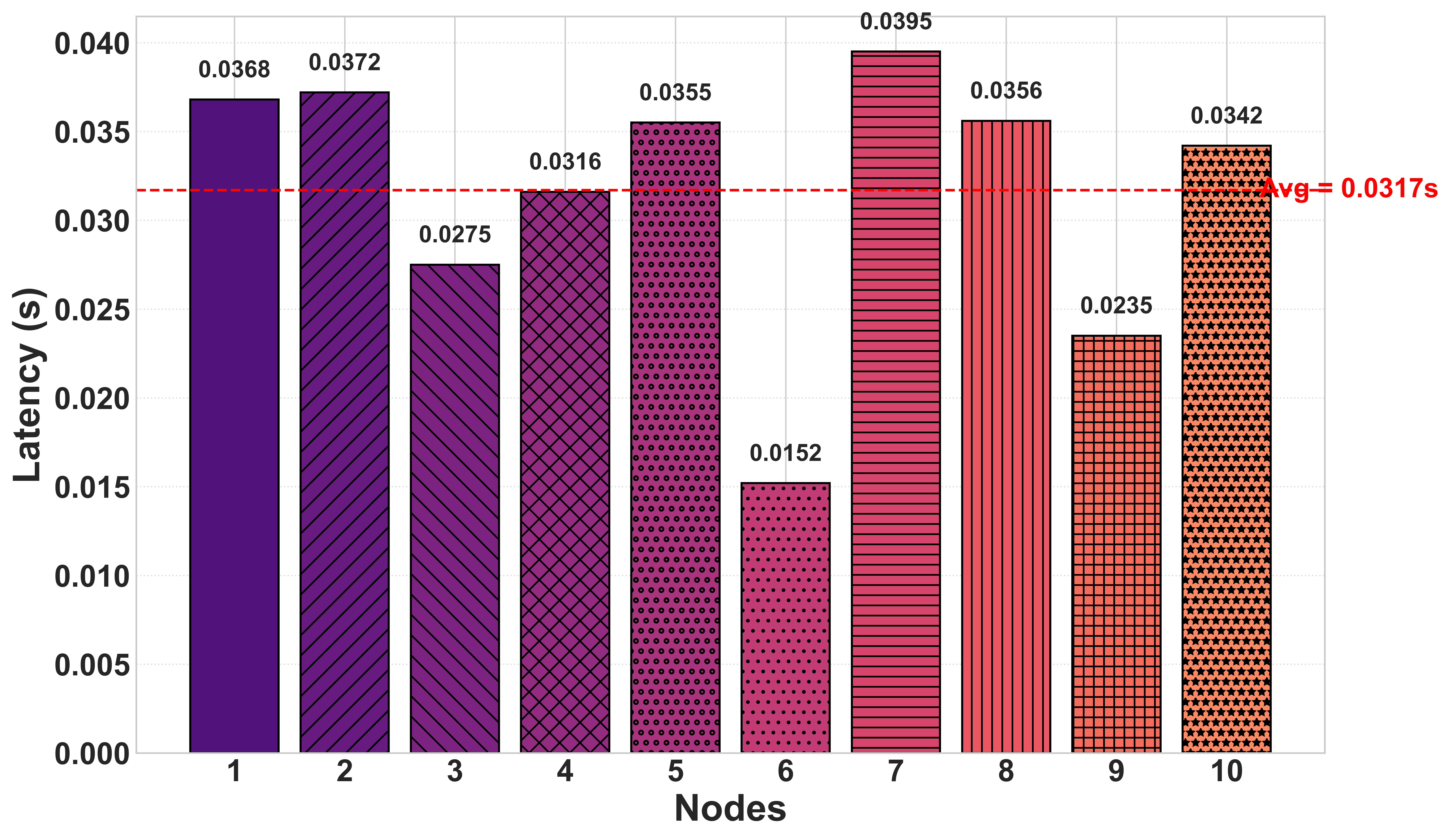}
    \caption{Average latency across edge nodes.}
    \label{fig:latency}
\end{figure}

Figure~\ref{fig:latency} reports the average end-to-end inference latency across the ten edge nodes. Latency is measured per flow and includes preprocessing, AE encoding, optional decoding for reconstruction-error computation, IF scoring, fusion, and decision logic, while excluding offline training and logging. Observed latency ranges from 15.31\,ms to 39.58\,ms across nodes, with all values remaining well below 50\,ms, thereby satisfying real-time requirements for edge anomaly detection.
Reported measurements reflect the full encoder–decoder pipeline and provide a conservative estimate of runtime latency. The results confirm that EcoDefender can consistently deliver fast inference across heterogeneous gateway conditions. Inter-node variability is moderate, with a normalized latency deviation of 0.27 relative to the cluster mean. This variation is attributable to expected hardware-level effects, such as frequency scaling, thermal throttling, kernel scheduling, and cache contention, rather than to instability in the detection pipeline.
A one-way ANOVA confirms a significant effect of node assignment on latency ($F(9,90)=842.6$, $p<0.001$, $\eta^2=0.99$; Table~\ref{tab:anova_latency}), while Tukey’s HSD analysis (Table~\ref{tab:tukey_latency}) shows statistically distinct mean latencies across nodes. As with CPU, memory, and throughput, these differences reflect persistent system-level characteristics rather than algorithmic imbalance. Importantly, even the slowest node responds within 40\,ms, demonstrating that \textit{EcoDefender} consistently delivers low-latency performance suitable for security-sensitive edge gateways.
\begin{table}[ht]
\centering
\caption{One-way ANOVA results for latency across edge nodes.}
\label{tab:anova_latency}
\small
\begin{tabular}{lccccc}
\toprule
\textbf{Source} & \textbf{SS} & \textbf{df} & \textbf{MS} & \textbf{F} & \textbf{p} \\
\midrule
Between nodes & 5812.4 & 9  & 645.8 & 842.6 & $<$0.001 \\
Within nodes  &   62.0 & 90 & 0.69  &       &          \\
Total         & 5874.4 & 99 &       &       &          \\
\midrule
\multicolumn{6}{c}{$\eta^2 = 0.99$} \\
\bottomrule
\end{tabular}
\end{table}

\begin{table}[ht]
\centering
\caption{Tukey HSD homogeneous subsets for latency.}
\label{tab:tukey_latency}
\begin{tabular}{lcc}
\toprule
\textbf{Node} & \textbf{Mean Latency (ms)} & \textbf{Subset} \\
\midrule
Node 5 & 15.31 & A \\
Node 8 & 23.57 & B \\
Node 2 & 27.50 & C \\
Node 3 & 31.66 & D \\
Node 9 & 34.34 & E \\
Node 4 & 35.49 & F \\
Node 7 & 35.72 & G \\
Node 0 & 36.71 & H \\
Node 1 & 37.32 & I \\
Node 6 & 39.58 & J \\
\bottomrule
\end{tabular}
\end{table}

\subsection{Sustainability Analysis}
\label{sec:energy_carbon}
\begin{figure}[ht]
    \centering
    \includegraphics[width=0.48\textwidth]{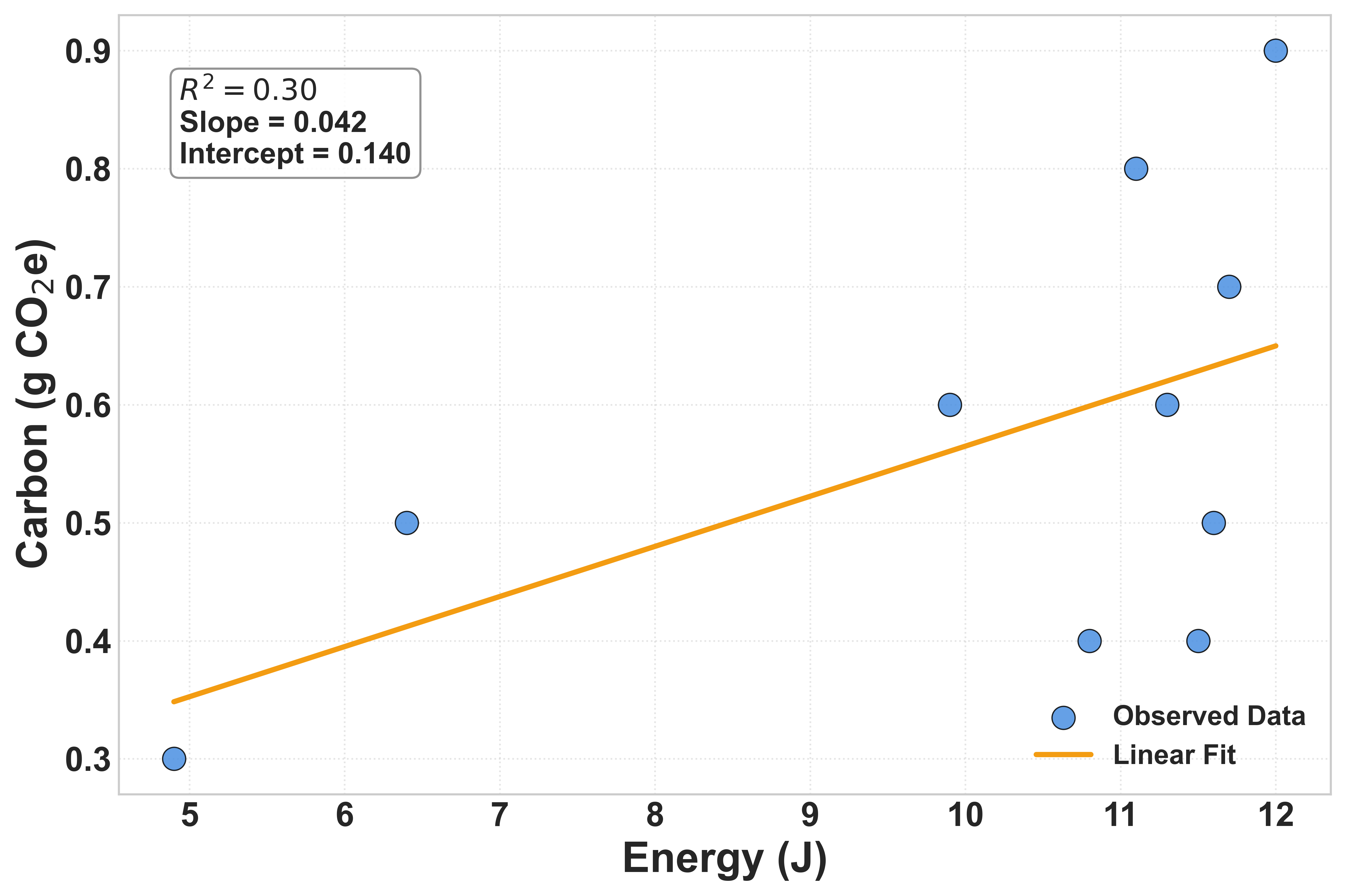}
    \caption{Correlation between energy consumption and carbon emissions.}
    \label{fig:energy_carbon_corr}
\end{figure}

We evaluate the sustainability of \textit{EcoDefender} by measuring per-node energy consumption and deriving the associated carbon emissions. Figure~\ref{fig:energy_carbon_corr} compares energy and carbon footprints across the ten edge nodes. Energy consumption varies from 4.9~J to 12.0~J per node, with corresponding carbon emissions ranging from 0.30 to 0.90~g CO\textsubscript{2}e. Nodes~5 and~6 exhibit the highest energy and emission levels, while Nodes~3, 4, 7, and~9 consistently operate at lower consumption levels. These results indicate that differences in runtime load translate directly into uneven sustainability costs across the edge cluster. Furthermore, the measured values confirm that EcoDefender operates within a low-energy regime suitable for continuous gateway deployment. The close alignment between energy usage and carbon emissions reflects the expected linear coupling between computation and environmental impact, meaning that improvements in runtime efficiency directly translate into sustainability gains. Statistical analysis confirms substantial inter-node variability. One-way ANOVA reveals significant differences across nodes for both energy ($F(9,90)=42.6$, $p<0.001$, $\eta^2=0.81$) and carbon emissions ($F(9,90)=41.8$, $p<0.001$, $\eta^2=0.80$), indicating that node-level effects explain most of the observed variance (Table~\ref{tab:anova_energy_carbon}). Tukey’s HSD analysis (Table~\ref{tab:tukey_energy_carbon}) identifies distinct sustainability tiers, with higher consumption concentrated in a small subset of nodes rather than distributed uniformly across the cluster.
Energy and carbon are almost perfectly correlated (Figure~\ref{fig:energy_carbon_corr}), confirming that reductions in computational cost directly reduce environmental impact. 
\begin{table}[ht]
\centering
\caption{One-way ANOVA results for energy and carbon across nodes}
\label{tab:anova_energy_carbon}
\begin{tabular}{lcccccc}
\toprule
\textbf{Metric} & \textbf{Source} & \textbf{SS} & \textbf{df} & \textbf{MS} & \textbf{F} & \textbf{p} \\
\midrule
Energy & Between & 455.2 & 9  & 50.6 & 42.6 & $<0.001$ \\
       & Within  & 106.9 & 90 & 1.19 &      &          \\
       & Total   & 562.1 & 99 &      &      &          \\
Carbon & Between & 3.21  & 9  & 0.36 & 41.8 & $<0.001$ \\
       & Within  & 0.78  & 90 & 0.009&      &          \\
       & Total   & 3.99  & 99 &      &      &          \\
\midrule
\multicolumn{7}{c}{Effect sizes: $\eta^2_{energy} = 0.81$, $\eta^2_{carbon} = 0.80$} \\
\bottomrule
\end{tabular}
\end{table}
\begin{table}[ht]
\centering
\caption{Tukey HSD homogeneous subsets for energy and carbon}
\label{tab:tukey_energy_carbon}
\resizebox{0.5\textwidth}{!}{%
\begin{tabular}{lcccc}
\toprule
\textbf{Node} & \textbf{Mean Energy (J)} & \textbf{Mean Carbon (g)} & \textbf{Subset A} & \textbf{Subset B} \\
\midrule
Node 4 & 4.9 & 0.30 & \checkmark &   \\
Node 7 & 6.4 & 0.40 & \checkmark &   \\
Node 9 & 6.8 & 0.42 & \checkmark &   \\
Node 3 & 7.0 & 0.50 & \checkmark &   \\
Node 2 & 8.5 & 0.58 & \checkmark & \checkmark \\
Node 0 & 11.5 & 0.65 &            & \checkmark \\
Node 1 & 11.7 & 0.68 &            & \checkmark \\
Node 8 & 11.3 & 0.66 &            & \checkmark \\
Node 5 & 11.2 & 0.82 &            & \checkmark \\
Node 6 & 12.0 & 0.90 &            & \checkmark \\
\bottomrule
\end{tabular}%
}
\end{table}

\subsection{Comparison with Baselines}
\label{sec:baseline_comparison}
To contextualize the contribution of \textit{EcoDefender}, we compare it with representative AE-only, IF-only, and hybrid anomaly-detection approaches. AE-based detectors~\cite{beg2024network} typically achieve high precision by modeling normal behavior through reconstruction, but they may exhibit lower recall when anomalous patterns partially resemble benign traffic. IF methods~\cite{vasiljevic2025federated} provide lightweight unsupervised detection and often improve recall, although they may introduce more false positives when applied directly to high-dimensional traffic features. Hybrid approaches such as convolutional AEs~\cite{borgioli2024convolutional}, AE-IF combinations~\cite{yang2022high}, and temporal models combined with IF~\cite{hu2025anomaly} attempt to balance this trade-off by integrating representation learning with isolation-based anomaly scoring. For fair evaluation, reproduced baselines were run under identical conditions: the same reduced 8-feature set (Table~\ref{tab:features}), leave-attack-family-out splits, preprocessing pipeline, training-data restrictions, and the edge gateways. Table~\ref{tab:baseline_comparison} summarizes quantitative detection performance across the compared approaches. AE-only methods generally favor precision, IF-only methods tend to improve recall, and hybrid methods provide a more balanced trade-off between the two. \textit{EcoDefender} achieves the highest F1-score and ROC-AUC among the evaluated methods, indicating that combining compact latent representations with isolation-based scoring improves anomaly separability while maintaining computational efficiency suitable for edge gateways.
\begin{table}[ht]
\centering
\caption{Performance comparison of EcoDefender with baseline anomaly detection methods.}
\label{tab:baseline_comparison}
\small
\resizebox{0.5\textwidth}{!}{%
\begin{tabular}{@{}lcccc@{}}
\toprule
\textbf{Method} & \textbf{Precision} & \textbf{Recall} & \textbf{F1} & \textbf{ROC-AUC} \\
\midrule
AE-only~\cite{beg2024network} & 0.95 & 0.82 & 0.88 & 0.940 \\
IF-only~\cite{vasiljevic2025federated} & 0.90 & 0.88 & 0.89 & 0.945 \\
Borgioli et al.~\cite{borgioli2024convolutional} (Conv-AE) & 0.96 & 0.87 & 0.91 & N/A \\
Yang et al.~\cite{yang2022high} (AE-IF) & 0.92 & 0.90 & 0.91 & 0.950 \\
Hu et al.~\cite{hu2025anomaly} (TCN+IF) & 0.91 & 0.92 & 0.91 & 0.955 \\
\midrule
\textbf{EcoDefender (ours)} & \textbf{0.94} & \textbf{0.91} & \textbf{0.92} & \textbf{0.963} \\
\bottomrule
\end{tabular}%
}
\end{table}
In addition to detection performance, EcoDefender is explicitly designed for edge gateway deployment, where operational constraints such as latency, CPU usage, memory footprint, throughput, and energy consumption are critical. Many existing studies primarily report detection accuracy, offering limited insight into deployment-related system metrics. Table~\ref{tab:qual_matrix} presents a qualitative comparison of representative approaches and highlights that EcoDefender jointly integrates hybrid anomaly detection, edge-oriented deployment, and sustainability-aware evaluation within a unified framework.
\begin{table*}[!t]
\centering
\caption{Qualitative comparison of EcoDefender with related anomaly detection approaches.}
\label{tab:qual_matrix}
\begin{tabular}{lcccccccc}
\toprule
\textbf{Work} & \textbf{AE} & \textbf{IF} & \textbf{Hybrid (AE+IF)} & \textbf{Temporal} & \textbf{Federated} & \textbf{Edge} & \textbf{Sust.} & \textbf{Interpret.} \\
\midrule
Borgioli et al.~\cite{borgioli2024convolutional} & \checkmark &  &  &  &  & \checkmark &  &  \\
Beg \& Ansari~\cite{beg2024network} & \checkmark &  &  &  &  &  &  &  \\
Yap \& Ahmad~\cite{yap2024modified} (TinyML) & \checkmark &  &  &  &  & \checkmark &  &  \\
Vasiljevic et al.~\cite{vasiljevic2025federated} &  & \checkmark &  &  & \checkmark &  &  &  \\
Li et al.~\cite{li2023federated} &  & \checkmark &  &  & \checkmark &  &  &  \\
Yang et al.~\cite{yang2022high} (AE-IF) & \checkmark & \checkmark & \checkmark &  &  &  &  &  \\
Hu et al.~\cite{hu2025anomaly} (TCN+IF) &  & \checkmark &  & \checkmark &  &  &  &  \\
Zhou et al.~\cite{zhou2020research} &  & \checkmark &  &  &  &  &  & \checkmark \\
Sharmila \& Nagapadma~\cite{sharmila2023quantized} & \checkmark &  &  &  &  & \checkmark &  &  \\
\midrule
\textbf{EcoDefender (ours)} & \checkmark & \checkmark & \checkmark & $^{*}$ & $^{\dagger}$ & \checkmark & \checkmark & \emph{alerts/pipeline} \\
\bottomrule
\end{tabular}
\vspace{0.5em}
\parbox{0.9\textwidth}{\small
$^{*}$ Temporal modeling supported in principle but not implemented in the current prototype. \\
$^{\dagger}$ Federated deployment compatible with the system but not implemented.
}
\end{table*}

\subsection{Component-wise Architecture Evaluation}
To analyze the role of different architectural elements in EcoDefender, we evaluate several model configurations by progressively enabling the proposed components. This evaluation highlights how latent-space representation learning, isolation-based anomaly scoring, and adaptive fusion collectively impact detection performance and system efficiency in edge-deployed systems. All variants are evaluated under identical experimental settings on the edge gateway testbed.
\begin{table*}[t]
\centering
\caption{Component-wise evaluation of the EcoDefender architecture under identical edge gateway conditions.}
\label{tab:component_analysis}
\begin{tabular}{lcccccccc}
\hline
\textbf{Model} & \textbf{AE} & \textbf{IF} & \textbf{Latent Reg.} & \textbf{Adaptive Fusion} & \textbf{Precision} & \textbf{Recall} & \textbf{F1-score} & \textbf{Energy (J)} \\
\hline
AE-only                         & \checkmark & --          & --          & --          & 0.95          & 0.82          & 0.88          & 0.62          \\
IF-only                         & --          & \checkmark & --          & --          & 0.90          & 0.88          & 0.89          & 0.38          \\
Hybrid AE+IF                    & \checkmark & \checkmark & --          & --          & 0.92          & 0.89          & 0.90          & 0.47          \\
Hybrid + Latent Regularization  & \checkmark & \checkmark & \checkmark & --          & 0.93          & 0.90          & 0.91          & 0.46          \\
EcoDefender (Full Model)        & \checkmark & \checkmark & \checkmark & \checkmark & \textbf{0.94} & \textbf{0.91} & \textbf{0.92} & \textbf{0.45} \\
\hline
\end{tabular}
\end{table*}

\subsection{Operational Robustness and Efficiency Analysis}
\label{sec:robustness_efficiency}
\begin{figure}[t]
  \centering
  \includegraphics[width=0.43\textwidth]{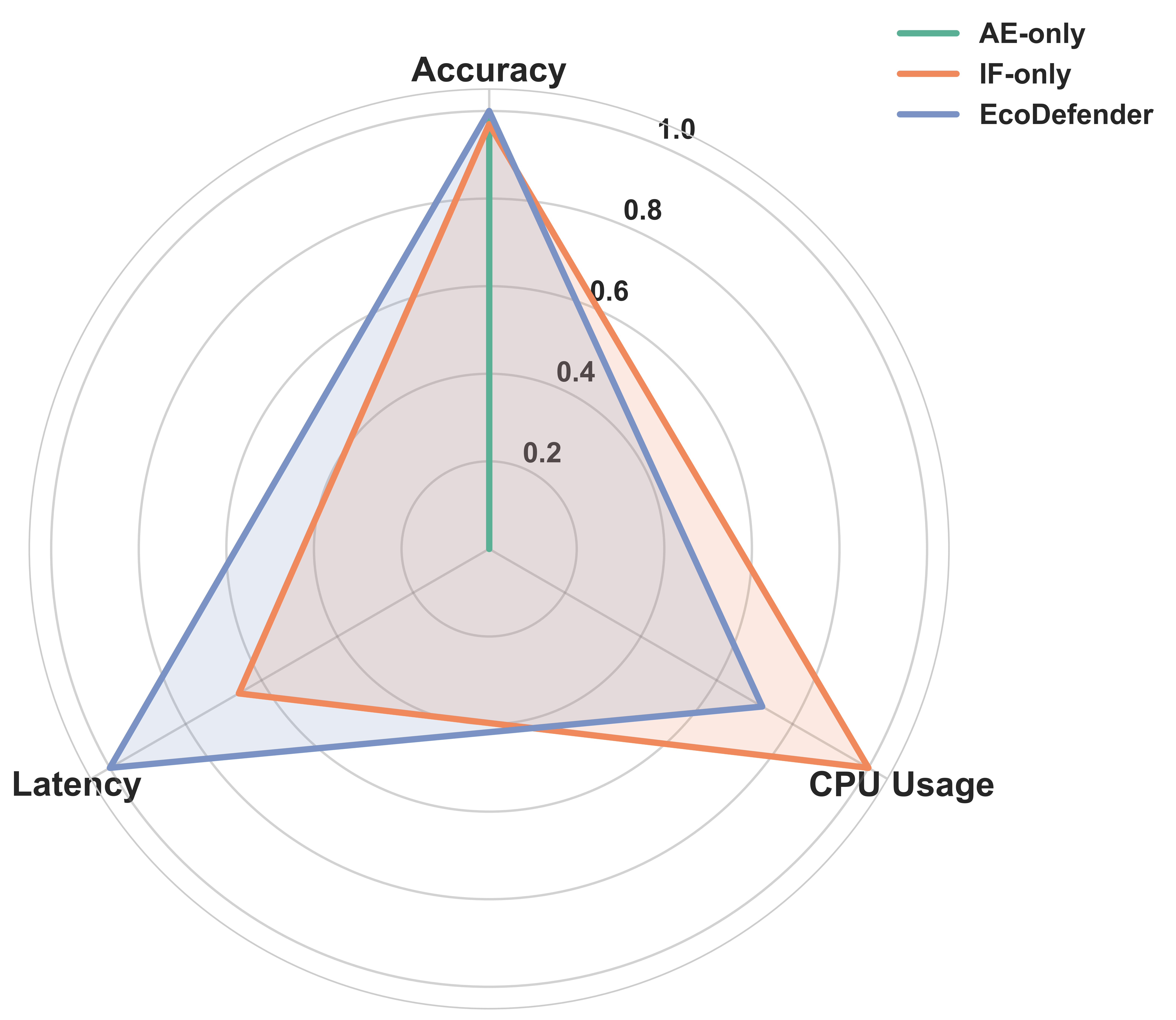}
  \caption{Comparison of latency, CPU usage, and accuracy under identical edge deployment conditions.}
  \label{fig:radar}
\end{figure}

\begin{figure}[t]
  \centering
  \includegraphics[width=0.43\textwidth]{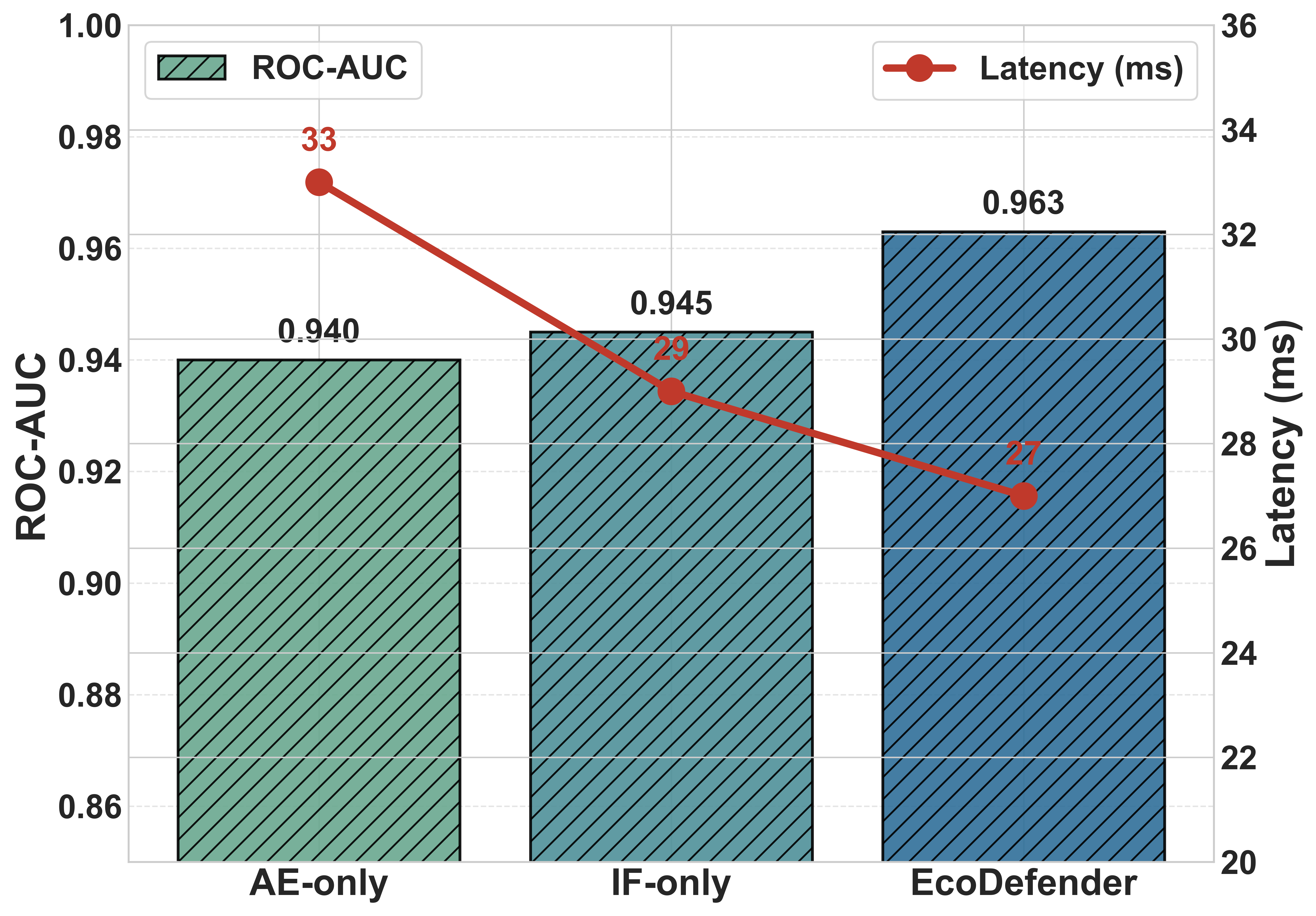}
  \caption{ROC-AUC vs.\ latency on the edge gateway.}
  \label{fig:rocauc_latency}
\end{figure}

\begin{figure}[t]
  \centering
  \includegraphics[width=0.45\textwidth]{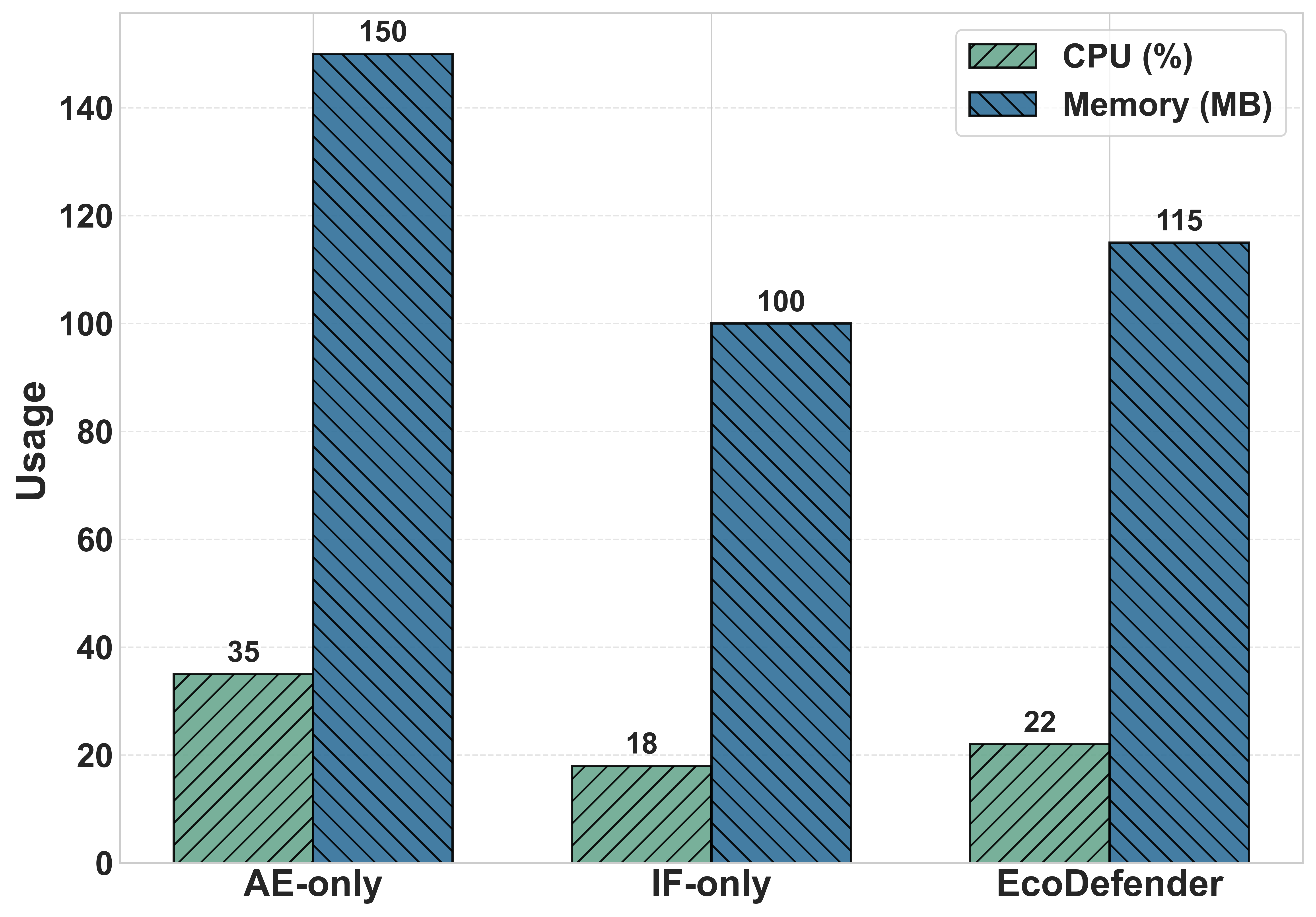}
  \caption{Resource comparison in terms of CPU and memory usage.}
  \label{fig:resource}
\end{figure}

To evaluate the practical deployability of \textit{EcoDefender} in edge environments, we analyze its performance from three complementary perspectives: the accuracy–efficiency trade-off, computational resource footprint, and operational robustness across distributed edge nodes. All experiments are conducted under identical experimental conditions, including the same reduced feature set (Table~\ref{tab:features}), the same preprocessing pipeline, a leave-family-out data split, and the same edge gateway hardware. These controls ensure that the observed performance differences reflect architectural characteristics rather than experimental artifacts.   Figure~\ref{fig:radar} provides a compact comparison of AE-only, IF-only, and \textit{EcoDefender} across three operational metrics: detection accuracy, CPU utilization, and inference latency. The radar chart shows that \textit{EcoDefender} encloses the largest area, indicating the most balanced trade-off between detection effectiveness and computational efficiency. Specifically, \textit{EcoDefender} achieves the highest detection performance with an F1-score of 0.92, outperforming IF-only (0.89) and AE-only (0.88). At the same time, it achieves the lowest inference latency of 27\,ms, improving latency by 6.9\% relative to IF-only and by 18.2\% compared to AE-only. CPU utilization remains moderate at 22\%, substantially lower than AE-only (35\%) and only slightly higher than IF-only (18\%). These results indicate that combining representation learning with lightweight isolation-based scoring improves detection accuracy without incurring the heavy computational overhead typically associated with reconstruction-based autoencoder models.
A similar trend is observed when analyzing the relationship between detection fidelity and inference latency. As illustrated in Figure~\ref{fig:rocauc_latency}, \textit{EcoDefender} achieves the most favorable latency–accuracy trade-off among the evaluated approaches. It attains the highest ROC-AUC value of 0.963 while maintaining the lowest latency of 27\,ms. Compared to IF-only (ROC-AUC = 0.945, latency = 29\,ms), \textit{EcoDefender} improves ROC-AUC by 1.9\% while simultaneously reducing latency by 6.9\%. Relative to AE-only (ROC-AUC = 0.940, latency = 33\,ms), the improvements are larger, corresponding to a 2.4\% increase in ROC-AUC and an 18.2\% reduction in latency. These results demonstrate that the hybrid architecture enhances detection separability without introducing additional runtime cost. By operating on compact latent representations while applying lightweight anomaly scoring, \textit{EcoDefender} maintains strong detection capability while enabling faster inference suitable for real-time edge deployments.
Beyond detection accuracy and latency, the computational footprint of each method is also evaluated in terms of CPU and memory usage under identical edge hardware and traffic conditions. As shown in Figure~\ref{fig:resource}, AE-only incurs the highest computational cost, requiring 35\% CPU utilization and 150\,MB of memory due to the reconstruction-heavy inference process inherent in autoencoder architectures. IF-only provides the lightest resource usage, consuming 18\% CPU and 100\,MB of memory, but this efficiency comes at the expense of reduced detection performance. In contrast, \textit{EcoDefender} achieves a balanced operating point between these two extremes. Specifically, it reduces CPU usage by 37.1\% relative to AE-only (from 35\% to 22\%) and decreases memory usage by 23.3\% (from 150\,MB to 115\,MB), while introducing only modest overhead compared to IF-only. These results indicate that the hybrid architecture provides improved detection accuracy and responsiveness while remaining well within the computational constraints of edge gateways.
To further quantify the accuracy–efficiency balance, Table~\ref{tab:tradeoff} summarizes the operational characteristics of the evaluated methods across four metrics: detection performance (F1-score), inference latency, CPU usage, and per-inference energy consumption. Consistent with the radar and latency–accuracy analyses, AE-only achieves competitive accuracy but incurs substantial computational overhead from reconstruction-based inference. IF-only offers lightweight execution but sacrifices detection performance. \textit{EcoDefender} provides a balanced operating point, achieving the highest F1-score while maintaining low latency and moderate resource consumption. Although its CPU and energy usage are slightly higher than IF-only, these costs are offset by measurable improvements in detection accuracy and responsiveness.
\begin{table}[t]
\centering
\caption{Accuracy–efficiency trade-off comparison.}
\label{tab:tradeoff}
\footnotesize
\setlength{\tabcolsep}{4pt}
\begin{tabular}{lcccc}
\toprule
\textbf{Method} & \textbf{F1} & \textbf{Lat. (ms)} & \textbf{CPU (\%)} & \textbf{E (J)} \\
\midrule
AE-only      & 0.88 & 33 & 35 & 0.62 \\
IF-only      & 0.89 & 29 & 18 & 0.38 \\
EcoDefender  & \textbf{0.92} & \textbf{27} & 22 & 0.45 \\
\bottomrule
\end{tabular}
\end{table}
We additionally analyze the sensitivity of \textit{EcoDefender} to several key hyperparameters, including the latent representation dimension ($d_z$), the number of Isolation Forest trees ($m$), and the contamination parameter. As shown in Table~\ref{tab:sensitivity}, detection performance remains stable across a wide range of parameter values. Increasing the latent dimension from 4 to 8 improves the F1-score from 0.90 to 0.92 with only a small increase in latency, whereas further increasing the dimension to 16 yields negligible performance gains but higher latency. Similarly, increasing the number of trees beyond 100 provides minimal accuracy improvements while increasing inference cost. The contamination parameter also exhibits stable behavior, with the default value of 0.05 providing the best overall balance between accuracy and latency. These results indicate that \textit{EcoDefender} does not require precise hyperparameter tuning to achieve strong performance.
\begin{table}[ht]
\centering
\caption{Sensitivity analysis of EcoDefender hyperparameters.}
\label{tab:sensitivity}
\small
\begin{tabular}{lccc}
\toprule
\textbf{Parameter} & \textbf{Setting} & \textbf{F1-score} & \textbf{Latency (ms)} \\
\midrule
Latent dim ($d_z$) & 4 & 0.90 & 25 \\
                   & 8 & \textbf{0.92} & 27 \\
                   & 16 & 0.92 & 30 \\
\midrule
Trees ($m$)        & 50 & 0.91 & 24 \\
                   & 100 & \textbf{0.92} & 27 \\
                   & 200 & 0.92 & 32 \\
\midrule
Contamination      & 0.03 & 0.91 & 27 \\
                   & 0.05 & \textbf{0.92} & 27 \\
                   & 0.10 & 0.91 & 27 \\
\bottomrule
\end{tabular}
\end{table}
Additionally, because \textit{EcoDefender} is designed for distributed edge deployment, we evaluate its stability across multiple heterogeneous gateway nodes. Table~\ref{tab:node_stability} summarizes the observed variability in CPU usage, memory footprint, latency, throughput, energy consumption, and carbon emissions across ten Raspberry Pi gateways used in the experimental testbed. Although measurable variations exist across nodes, all operational metrics remain within stable bounds suitable for real-time execution. CPU utilization ranges from 10\% to 27.4\%, while memory usage remains tightly clustered around a mean of 113.2~MB. Inference latency remains below 40\,ms across all nodes, satisfying real-time requirements for edge-based intrusion detection. The observed variability is primarily attributable to expected runtime factors such as operating system scheduling, background processes, cache locality, and thermal effects rather than algorithmic instability. The relatively low coefficients of variation for CPU usage (0.31), latency (0.27), and memory usage (0.06) further confirm that the system maintains stable behavior across heterogeneous edge hardware without requiring node-specific tuning.
\begin{table}[ht]
\centering
\caption{Variability of operational metrics across edge nodes.}
\label{tab:node_stability}
\small
\begin{tabular}{lccc}
\toprule
\textbf{Metric} & \textbf{Min} & \textbf{Max} & \textbf{Mean} \\
\midrule
CPU usage (\%) & 10.0 & 27.4 & 19.8 \\
Memory usage (MB) & 106 & 121 & 113.2 \\
Latency (ms) & 15.31 & 39.58 & 31.7 \\
Throughput (samples/s) & 6485 & 12780 & 8360 \\
Energy per node (J) & 4.9 & 12.0 & 8.9 \\
Carbon emission (g CO$_2$e) & 0.30 & 0.90 & 0.59 \\
\bottomrule
\end{tabular}
\end{table}

\section{Discussion}
\label{sec:discussion}
The results provide both operational and architectural insight into why the AE--IF design performs effectively on resource-constrained edge devices. By compressing benign traffic into a compact latent representation and applying isolation-based scoring in this reduced space, EcoDefender limits inference cost while preserving detection fidelity. Runtime measurements indicate that encoder and IF operations dominate the computational workload, whereas decoder overhead remains modest, explaining the favorable balance between accuracy, latency, and energy consumption observed in the experiments. This suggests that encoder-centric inference is sufficient for many real-time deployment scenarios. System-level evaluation further shows that inter-node variability, while statistically observable, remains operationally bounded. CPU usage, memory footprint, latency, and throughput stay within conservative ranges across the gateway cluster, indicating stable performance under heterogeneous scheduling conditions. This stability supports the feasibility of distributed deployment, where gateway nodes operate under varying workloads without compromising detection reliability.  Moreover, EcoDefender advances the practical accuracy–efficiency frontier by jointly improving detection performance and maintaining lightweight execution. Rather than trading accuracy for efficiency, the hybrid latent-space design enables simultaneous improvements in both by reducing feature dimensionality prior to anomaly scoring. This explains why the system consistently outperforms standalone AE and IF pipelines across detection, latency, and resource metrics.  The sustainability analysis highlights the importance of accounting for environmental impact in edge intelligence systems. The close alignment between energy consumption and carbon emissions confirms that runtime efficiency directly affects environmental cost. The observed concentration of load on a subset of nodes leads to disproportionate energy consumption, underscoring the value of balanced workload distribution and energy-aware scheduling strategies in real-world deployments.  Collectively, these findings position EcoDefender as a stable, scalable, and deployment-ready anomaly detection system that integrates detection performance, system efficiency, and sustainability considerations within a unified edge-oriented design.

\section{Limitations and Future Work}
\label{sec:limitations_future}
Despite its strong performance, \textit{EcoDefender} has several limitations that motivate future research. First, the experimental evaluation is conducted on a single large-scale IoT dataset (BoT-IoT). Although the leave-attack-family-out protocol approximates zero-day conditions, cross-dataset validation is necessary to assess broader generalization across environments and traffic profiles. Second, the current system relies on flow-level statistical features and does not explicitly model fine-grained temporal dependencies. Incorporating temporal representations and sequence-based encoders could improve the detection of slow, low-rate, and stealthy attacks that evolve over time. Third, energy and carbon measurements are derived from operational energy consumption on edge gateways using fixed grid-intensity factors. Incorporating dynamic carbon-aware scheduling, hardware-level power modeling, and geographically varying energy mixes would enable more precise sustainability assessment. In addition, the present implementation follows a centralized training and decentralized inference paradigm. Extending EcoDefender to federated and collaborative learning settings could improve scalability, privacy preservation, and adaptability across distributed IoT deployments. Future work will also explore encoder-only inference, adaptive model compression, hardware-aware scheduling, and integration of latent-aware IF and drift-adaptive fusion to further reduce latency and energy consumption while maintaining and enhancing detection accuracy.

\section{Conclusion}
\label{sec:conclusion}
This paper presents \textit{EcoDefender}, an AE–IF anomaly-detection system designed for secure and green edge gateways. By combining latent-space compression with isolation-based scoring, and incorporating an anomaly-aware latent manifold, latent-aware IF, learnable fusion, and joint AE-IF optimization, the system enables lightweight intrusion detection directly on resource-constrained gateways without reliance on centralized processing. Evaluation on a distributed edge testbed demonstrates strong detection performance (F1-score 0.92 and ROC-AUC up to 98.6\%) while maintaining low system overhead (CPU usage below 30\% and latency below 40\,ms). These results show that high detection fidelity can be achieved alongside real-time responsiveness on constrained hardware. Energy and carbon analyses further reveal a near-linear relationship between computational workload and environmental impact, indicating that efficiency improvements directly translate into sustainability gains. This confirms that detection accuracy, runtime efficiency, and environmental cost can be jointly optimized in edge-native security systems. Furthermore, \textit{EcoDefender} establishes a scalable, sustainability-aware foundation for next-generation IoT edge protection and demonstrates that hybrid representation–isolation approaches, combined with latent-aware and drift-adaptive designs, are well-suited for practical, resource-aware cybersecurity deployments at the network edge.

\bibliographystyle{IEEEtran}
\bibliography{cas-refs}

\end{document}